\documentclass[sigconf]{acmart}

\usepackage{threeparttable}
\usepackage[utf8]{inputenc} % alzlow utf-8 input

\usepackage{booktabs}       % professional-quality tables
\usepackage{nicefrac}       % compact symbols for 1/2, etc.
\usepackage{microtype}      % microtypography
\usepackage{xcolor}         % colors
\usepackage{soul}
\usepackage{url}
\usepackage[font={small,bf}]{caption}
\usepackage{graphicx}
\usepackage{amsmath}
\usepackage{amsthm}
\usepackage{algorithm}
\usepackage{algorithmic}
\usepackage{multirow}
\usepackage{subcaption}

\usepackage{bm}
\usepackage{color}

\usepackage{tablefootnote}
\usepackage{lipsum}
\usepackage{float} 
\usepackage{tcolorbox}

\usepackage{xspace}
\usepackage{stfloats}
\usepackage{marvosym}
\usepackage{enumitem}
\usepackage{pifont}%http://ctan.org/pkg/pifont

\usepackage{color, colortbl}

\AtBeginDocument{%
  \providecommand\BibTeX{{%
    \normalfont B\kern-0.5em{\scshape i\kern-0.25em b}\kern-0.8em\TeX}}}

\setcopyright{acmcopyright}
\copyrightyear{2025}
\acmYear{2025}
\acmDOI{XXXXXXX.XXXXXXX}

\acmConference[Conference acronym 'XX]{Make sure to enter the correct
  conference title from your rights confirmation emai}{June 03--05,
  2018}{Woodstock, NY}

\acmPrice{15.00}
\acmISBN{978-1-4503-XXXX-X/18/06}

\begin{document}
% \maketitle
\title{Enhancing CTR Prediction with De-correlated Expert Networks}

% \author{Jiancheng Wang$^{1}$, Mingjia Yin$^{1}$, 
%  Hao Wang$^1\dagger$, Enhong Chen$^1$}
% \affiliation{%
%     \institution{$^1$ University of Science and Technology of China \& State Key Laboratory of Cognitive Intelligence}
%     \country{}
% }
% \email{{wangjc830, mingjia-yin}@mail.ustc.edu.cn,{wanghao3, cheneh}@ustc.edu.cn}
\author{Jiancheng Wang, Mingjia Yin, 
 Hao Wang$\dagger$, Enhong Chen}
\affiliation{%
    \institution{University of Science and Technology of China \& State Key Laboratory of Cognitive Intelligence}
    \country{}
}
\email{{wangjc830, mingjia-yin}@mail.ustc.edu.cn,{wanghao3, cheneh}@ustc.edu.cn}

\renewcommand{\shortauthors}{Jiancheng, et al.}

\begin{abstract}
    Modeling feature interactions is essential for accurate click-through rate (CTR) prediction in advertising systems. 
    Recent studies have adopted the Mixture-of-Experts (MoE) approach to improve performance by ensembling multiple feature interaction experts. 
    These studies employ various strategies, such as learning independent embedding tables for each expert or utilizing heterogeneous expert architectures, to differentiate the experts, which we refer to expert \emph{de-correlation}. 
    However, it remains unclear whether these strategies effectively achieve de-correlated experts. 
    To address this, we propose a De-Correlated MoE (D-MoE) framework, which introduces a Cross-Expert De-Correlation loss to minimize expert correlations.
    Additionally, we propose a novel metric, termed Cross-Expert Correlation, to quantitatively evaluate the expert de-correlation degree. 
    Based on this metric, we identify a key finding for MoE framework design: \emph{different de-correlation strategies are mutually compatible, and progressively employing them leads to reduced correlation and enhanced performance}.
    Extensive experiments have been conducted to validate the effectiveness of D-MoE and the de-correlation principle. 
    Moreover, online A/B testing on one of the largest advertising platforms demonstrates that D-MoE achieves a significant 1.19\% Gross Merchandise Volume (GMV) lift compared to the Multi-Embedding MoE baseline.
    % Anonymous code is provided at \textcolor{blue}{\url{https://anonymous.4open.science/r/D-MoE/}}.
    % These results demonstrate the effectiveness of our proposed approach in improving the performance of advertising systems.
  % Modeling feature interactions play a crucial role in accurately predicting click-through rates (CTR) in advertising systems. 
  % Several works adopt the Mixture-of-Experts (MoE) to ensemble several feature interaction experts to boost the performance.
  % They employ various approaches to differentiate these experts by learning independent embedding tables for each expert or adopting heterogeneous expert architectures.
  % However, it's still a mystery whether they succeeded in learning different representations.
  % In this paper, we present a novel Cross-Expert Correlation metric and found that the representation of different experts are still strongly correlated with each other.
  % To this end, we propose a De-Correlated MoE, equipped with a Cross-Expert De-Correlation loss between the outputs of experts.
  % We conduct comprehensive evaluations on two popular public datasets, and our proposed MoE significantly outperforms several best-performing baselines.
  % During online A/B testing in one of the world's largest advertising platforms, D-MoE achieves 1.03\% cost lift and 1.19\% GMV lift over the base model.

\end{abstract}


\maketitle

\section{Introduction}
\begin{figure}[t]
    \centering
    \includegraphics[width=0.9\linewidth]{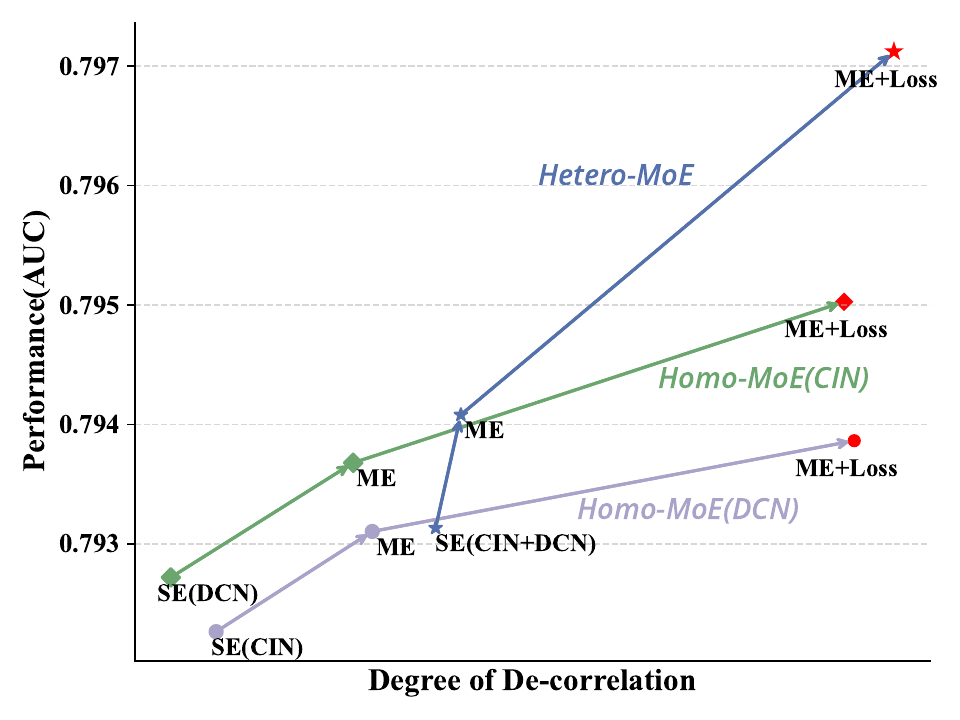} 
    \caption{
    An empirical study that reveals a strong positive connection between expert de-correlation and recommendation performance: higher de-correlation leads to better model performance.
    % Empirical study on the connection between MoE expert de-correlation (x-axis) and recommendation performance (y-axis).
    }
    \label{fig:intro}
\end{figure}

Click-through rate (CTR) estimation plays a core role for online advertising and recommender systems~\cite{DLRSSurvey2019, ctr_survey}.
CTR prediction involves multi-categorical feature data, where modeling feature interactions is critical for achieving accurate predictions.
Much efforts have been devoted to improving recommendation performance by designing sophisticated feature interaction models~\cite{DLRSSurvey2019}, ranging from explicit feature interaction models~\cite{FM2010, FwFM2018, FmFM2021, AFM2017, FFM2016} to complex DNN-based models~\cite{NFM2017, DeepFM2017, FibiNet2019, DCNv22021, PNN2016, xDeepFM2018, FinalMLP2023, zhu2023final, dcnv3, AutoInt2019, AutoAttention2022}.

Orthogonal to designing standalone sophisticated feature interaction modules, the Mixture-of-Experts (MoE) approach has been introduced to enhance recommendation performance.
Initially proposed by~\citet{jacobs1991adaptive}, MoE has garnered significant attention from both academia and industry~\cite{jacobs1991adaptive, eigen2013learning, shazeer2017outrageously, dai2024deepseekmoe}.
It leverages a combination of specialized models, known as \textit{experts}, to handle complex tasks more efficiently by dynamically selecting the most appropriate expert for different input data.
Specifically, MoE-based recommender systems can integrate multiple specialized feature interaction experts, thereby improving recommendation performance, especially in multi-task or multi-domain (or multi-scenario) recommendation~\cite{moe, moe_survey, MMoE2018, PLE2020, STEM2023, lin2024crocodile}.

Ideally, each expert in MoE is designed to specialize in distinct aspects of the data, necessitating a high degree of differentiability among them, which we refer to \emph{de-correlation} among experts.
Previous works have utilized different strategies to de-correlate experts intentionally or unintentionally, such as learning multiple independent embedding tables for each expert~\cite{guo2023multi-embedding} or employing heterogeneous experts~\cite{zhang2022dhen, pan2024ads}.
However, it remains unclear whether these approaches have successfully learned specialized experts.

To achieve improved expert de-correlation, we propose a De-Correlated MoE (D-MoE), which incorporates a cross-expert de-correlation loss between the expert outputs.
Compared to existing methods that seek expert de-correlation from a model design perspective, this loss function directly regards the correlation among experts as the optimization objective.
This straight-through strategy is highly flexible and model-agnostic, ensuring promising compatibility with other de-correlation strategies, including previously mentioned multi-embedding and heterogeneous experts.
Therefore, D-MoE can achieve better expert de-correlation, thereby leading to improved recommendation performance.

To conduct an in-depth investigation on the expert de-correlation problem, we introduce a novel metric termed \emph{Cross-Expert Correlation}, which measures the average dimension-wise Pearson correlation coefficient between outputs generated by different experts.
Based on this metric, we can quantitatively analyze existing MoE de-correlation strategies, as illustrated in Fig.~\ref{fig:intro}.
In the figure, we can derive these observations:
(1) starting from either a Single-Embedding homogeneous or heterogeneous MoE, further employing the multi-embedding paradigm and then the de-correlation loss can progressively de-correlate the experts and improve the performance;
(2) the heterogeneous MoE architecture consistently outperforms the homogeneous MoE counterpart, under the same setting, e.g., single-embedding, multi-embedding, or multi-embedding with the de-correlation loss.
These observations substantiate an interesting principle for MoE framework designs: various de-correlation strategies are compatible to each other, and progressively employing them leads to decreased correlation and better performance.
Our main contributions are summarized as follows:
% In the figure, all three de-correlation strategies can de-correlate the experts to some extent.
% Surprisingly, experts still exhibit strong correlations with each another, despite the use of meticulously designed strategies like multi-embedding or heterogeneous experts.
% Only equipped with the proposed de-correlation loss can these MoE frameworks achieve a desired degree of de-correlation.
% More importantly, the figure reveals a strong positive connection between the expert de-correlation and model performance, which can serve as a universal principle for MoE framework designs.

% We conducted extensive experiments to validate the superiority of the proposed D-MoE framework. 
% D-MoE significantly outperforms competitive baselines on two popular public datasets by a considerable margin. 
% Using the proposed metric, we quantitatively analyzed the de-correlation capabilities of various strategies, including the proposed CorrLoss, multi-embedding, and heterogeneous experts, providing strong empirical evidence for the de-correlation principle in MoE framework design. 
% Additionally, we carefully designed experiments to investigate the loss function, including its formulation and the layers to which the de-correlation loss is applied, offering valuable insights into MoE expert de-correlation. 
% Furthermore, during online A/B testing on one of the world's largest advertising platforms, our De-Correlated MoE achieved a remarkable 1.03\% increase in cost efficiency and a 1.19\% lift in GMV (Gross Merchandise Volume) compared to the base model.

\begin{itemize}[leftmargin=*]
    % \item To the best of our knowledge, we are the first to explicitly and quantitatively investigate the MoE de-correlation problem in recommender systems.
    \item We propose a De-Correlated MoE (D-MoE), which incorporates a Cross-Expert De-Correlation loss between expert outputs. This loss is highly flexible and model-agnostic, thereby exhibiting promising compatibility with other de-correlation strategies.
    \item We derive a novel metric termed Cross-Expert Correlation that can be used to quantitatively investigate the MoE expert de-correlation degree, based on which we find an interesting principle for MoE framework designs: various de-correlation strategies are compatible to each other, and progressively employing them leads to decreased correlation and better performance.
    \item Extensive experiments have been conducted to demonstrate the superiority of D-MoE, along with comprehensive verification of the loss design. Besides, the proposed de-correlation principle is further validated with convincing empirical evidences.
\end{itemize}

\paragraph{}

% \paragraph{Scalability of CTR models.} ~\cite{guo2023multi-embedding, zhang2024wukong, hstu2024} \textcolor{gray}{\lipsum[1]}

% \paragraph{MoE for Recommendation.} \textcolor{gray}{\lipsum[1]}
% Multi-Learning for Recommendation~\cite{MMoE2018, PLE2020, STEM2023, Cro}

% \paragraph{Correlation of Existing MoE.} \textcolor{gray}{\lipsum[1]}

% \paragraph{Our De-correlated MoE architecture.} \textcolor{gray}{\lipsum[1]}

% \paragraph{Evaluation.} \textcolor{gray}{\lipsum[1]}

% \paragraph{Contribution.} \textcolor{gray}{\lipsum[1]}

\section{Preliminaries}

\subsection{Problem Definition}
Click-Through Rates (CTR) prediction aims to predict the probability of a user clicking an advertisement in advertising systems, which is a classification problem.
A typical CTR prediction problem can formally defined as $\{ \mathcal{X}, \mathcal{Y} \}$, where $\mathcal{X}$ is a multi-categorical feature set, while $\mathcal{Y} \in \{0, 1\}$ is a label set indicating whether users have clicked the advertisement.
Usually, $\mathcal{X}$ includes user, item, and contextual features.
Then a recommendation model aims to learn a mapping from $\mathcal{X}$ to $\mathcal{Y}$.

\subsection{Mixture-of-Experts}
The Mixture of Experts (MoE) architecture has been widely adopted in recommendation systems, particularly for multi-task or multi-domain (or multi-scenario) recommendation tasks~\cite{moe, moe_survey, MMoE2018, PLE2020, STEM2023, lin2024crocodile}. 
The MoE frameworks usually consist of several key modules, including embedding, expert, gating and tower layer, along with a binary cross entropy loss function for optimization.

\begin{figure*}
    \centering
    \includegraphics[width=0.95\linewidth]{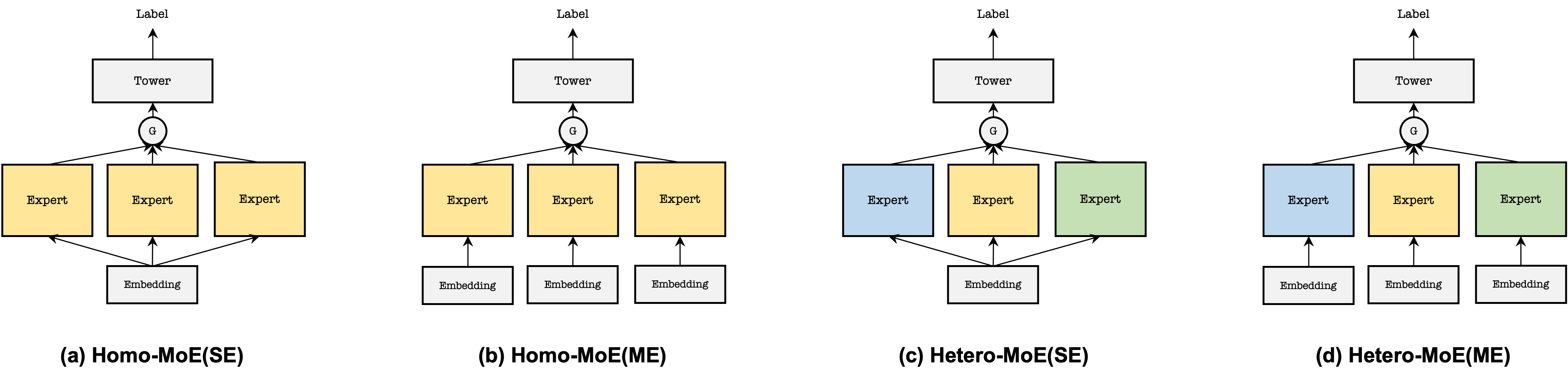}
    \vspace{-0.25cm}
    \caption{Different Variants of Mixture of Experts (MoE). Shared-Embedding (SE) doesn't  disentangle representations, whereas Multi-Embedding (ME) provides each expert with an independent embedding table. Homo-MoE consists of homogeneous experts with distinct parameters, while Hetero-MoE comprises heterogeneous experts to capture unique interactions.}
    \label{fig:moe_zoo}
\end{figure*}

% Following \cite{MMoE2018}, we employ a group of individual networks as experts instead of sharing bottom. In our work, experts can be homogeneous or heterogeneous.
% Inspired by \cite{STEM2023}, we employ shared-embedding(SE) and multi-embedding(ME). For SE, all experts share one embedding. For ME, each expert is equipped with an independent embedding table to de-correlate experts from input. In particular, when we input sparse feature denoted as $x$, $\mathbf{E}$ represents a set of embedding table. 
\subsubsection{\textbf{Embedding}}\label{sec:pre-moe-emb}

In CTR prediction, each feature $\bm{x}$ belongs to a feature field $F_i$ with cardinality $D_i$, where $D_i$ is the number of unique features in the feature field.
In typical MoE frameworks like MMoE \cite{MMoE2018} and PLE \cite{PLE2020}, each feature field $\bm{x}$ will be associated with an embedding lookup table $\bm{E} \in \mathbb{R}^{D_i \times d}$, where $d$ denotes the embedding dimension size.
This can be formally defined as:
\begin{equation}
    \bm{e}_i = \mathbf{E}(\bm{x}_i).
\end{equation}

These embedding tables of different feature fields will be shared among different experts defined in the next subsection, which we refer as a \textit{shared-embedding (SE)} paradigm.

\subsubsection{\textbf{Experts}}

Experts are the most important parts of a MoE framework, where each expert is supposed to specialize at distinct aspects of the data.
In the context of recommender systems, each expert is a feature interaction model, like FM~\cite{FM2010}, DeepFM~\cite{DeepFM2017} or DCNv2~\cite{DCNv22021}.
The formerly defined feature embeddings $\bm{e}$ will be fed into each expert to produce expert outputs:
\begin{align}
    \bm{o}_i^{(m)} &= \text{Expert}^{(m)}(\bm{e}_i), \quad \forall m \in \{1, 2, \ldots, M\},
\end{align}
where \( \bm{o}_i^{(m)} \) represents the output of the \( m \)-th expert, and $M$ is the number of experts.

The selection of each expert is arbitrary, provided that the expert can generate meaningful representations based on feature embeddings. 
Despite this flexibility, most existing MoE frameworks utilize identical feature interaction models for their experts~\cite{MMoE2018, PLE2020}, a design we refer to as \emph{homogeneous} expert design.

% These experts can be either homogeneous or heterogeneous, marking a departure from the traditional shared-bottom architectures that rely on a single shared network for all tasks.
% This flexibility allows for more specialized and potentially more effective feature processing.

\subsubsection{\textbf{Gating}} \label{gating}
After generating expert outputs, the MoE framework activates different experts depending on the input characteristics with a gating mechanism.
Specifically, it computes a set of expert weights, typically using a Softmax function, to assign probabilities to each expert based on the input's relevance to their learned knowledge.
The gating mechanism is supposed to ensure that only the most appropriate experts are activated, enabling specialized processing of the input.
This process can be formally defined as:
\begin{equation}
    \mathbf{h}_i=\sum_{m=1}^M g^{(m)}(\bm{x}_i) \cdot \bm{o}_i^{(m)},
\end{equation}
where $\sum_{m=1}^M g^{(m)}(\bm{x}_i)=1$. $g^{(m)}(\bm{x}_i)$ indicates the weight assigned to the expert output $\bm{o}_i^{(m)}$.

% In MoE, the gating mechanism is a critical component that determines how input data is routed to different experts.
% It computes a set of weights, typically using a softmax function, to assign probabilities to each expert based on the input's relevance to their specialization. 
% These weights ensure that only the most suitable experts are activated, enabling efficient and specialized processing of the input.

% After we obtain the output of experts $o_i,i=1,2,...,M$, gating mechanism integrates outputs from experts. The output of the $i$-th sample can be formulated as:
% \begin{equation}
%     \mathbf{h}_i=\sum_{m=1}^M g_i^{m} \cdot o_i^{m}
% \end{equation}
% where $\sum_{m=1}^M g_i^m=1$. $g_i^m$ indicates the probability to choose the expert output $o_i$.
% \paragraph{\todo{Concrete gating such as MMoE,PEPNet.}}

\subsubsection{\textbf{Towers and Loss Function}}
The gated expert outputs will be processed with a fusion tower to obtain the final output:
\begin{equation}
    \hat{y}_i = \sigma(\text{MLP}(\mathbf{h}_i)),
\end{equation}
where $\sigma$ is the sigmoid function. For CTR task, we often choose binary cross-entropy(BCE) loss function as the objective function. The final loss is formulated as follows:
\begin{equation}
\mathcal{L}=-\frac{1}{N}\sum_i^N y_i \log \left(\hat{y}_i\right)+\left(1-y_i\right) \log \left(1-\hat{y}_i\right)
\end{equation}
where $y_i$ is the ground truth of sample $i$.

\subsection{MoE Experts De-correlation} \label{2.3}

Ideally, each expert in MoE is designed to specialize in distinct aspects of the data, necessitating a high degree of differentiability among them, which we refer to \emph{de-correlation} among experts.
Previous works have utilized different strategies to de-correlate experts intentionally or unintentionally, such as learning multiple independent embedding tables for each expert~\cite{guo2023multi-embedding} or employing heterogeneous expert architectures~\cite{zhang2022dhen} as follows:

\paragraph{\textbf{Multi-embedding}}

Unlike the \textit{shared-embedding (SE)} paradigm mentioned in Sec.~\ref{sec:pre-moe-emb}, the Multi-Embedding(ME) paradigm, introduced in \cite{guo2023multi-embedding, pan2024ads}, addresses embedding collapse by learning distinct data patterns through specialized embedding tables. 
Furthermore, \cite{STEM2023, lin2024crocodile} have implemented multi-embedding within MoE architectures for multi-task learning, resolving insufficient embedding learning challenges while enhancing feature diversity.

This ME paradigm enables simultaneous learning of divergent data patterns and diverse feature interactions, thereby demonstrating intrinsic de-correlation at the input level.

\paragraph{\textbf{Heterogeneous MoE}}
The MoE architecture has been predominantly applied in multi-task recommendation systems.
While traditional approaches employ task-specific towers with shared-bottom networks \cite{SharedBottom1997}, recent advancements in MoE architectures have evolved from partially shared experts \cite{ESMM2018,Cross-Stitch,PLE2020} to mutually independent expert configurations \cite{STEM2023, lin2024crocodile, dai2024deepseekmoe}. 

In large-scale CTR predictions, ~\cite{zhang2022dhen, me_dhen} also emphasize that ensemble different interactions to leverage strengths of heterogeneous modules and learn a hierarchy of the interactions under different orders. 
We conclude these advancements demonstrates de-correlation at the expert level in enhancing model performance.

However, the studies above do not emphasize de-correlation. To address this, we provide an evaluation of the degree of de-correlation in Sec. \ref{method_analysis}.

\section{De-correlated MoE}

% We propose a multi-embedding Mixture of Experts architecture enhanced with a De-correlation Loss(DMoE) to diversify the outputs of each expert. 

\begin{figure*}[ht]
    \centering
    \includegraphics[width=0.6\linewidth]{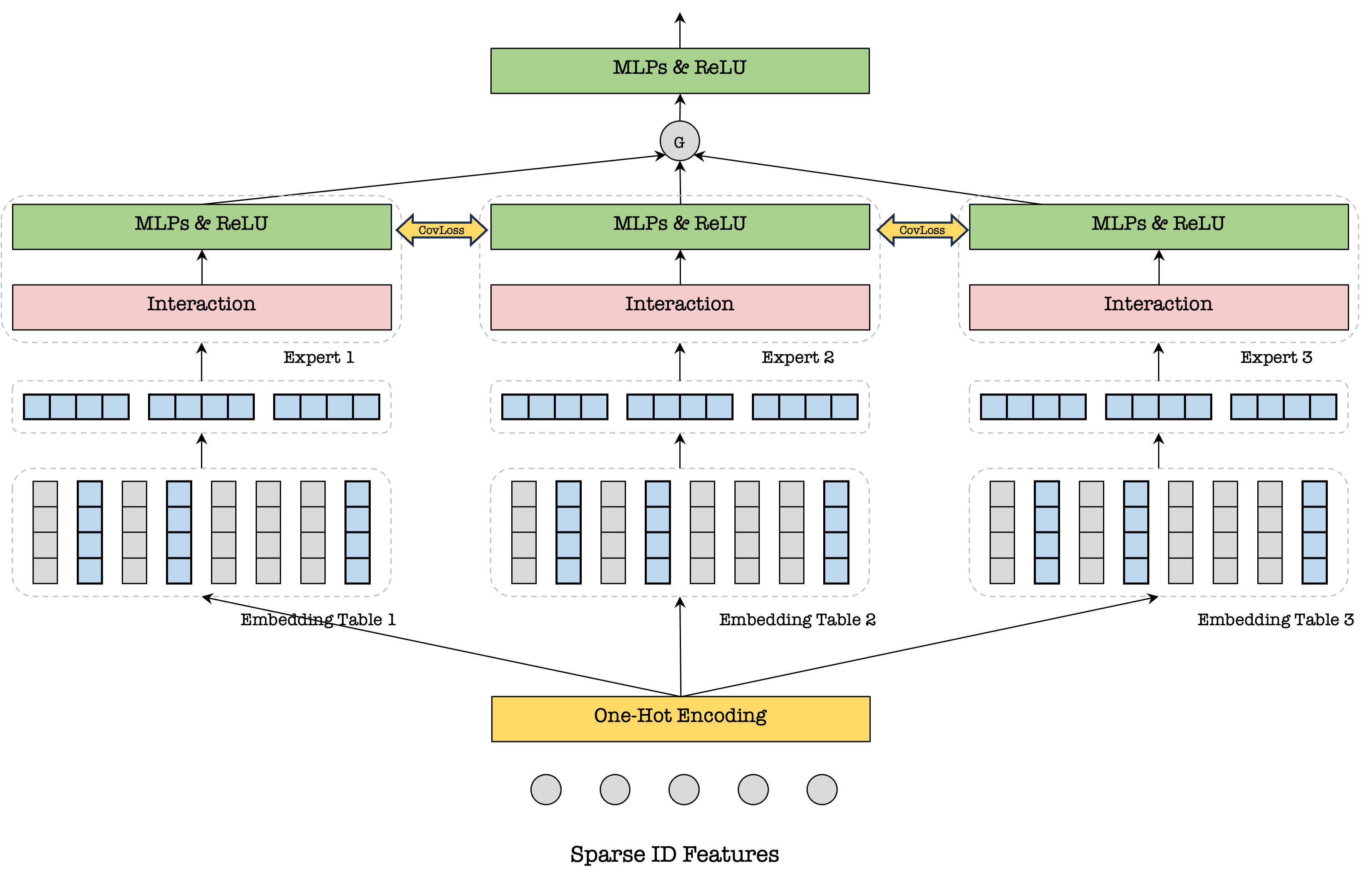}
    \vspace{-0.25cm}
    \caption{Architecture of the De-correlated Mixture-of-Experts. Each Interaction model represents an expert. We employ a de-correlation loss upon the outputs between expert pairs. Experts can be either homogeneous or heterogeneous. We use MLPs \& ReLU layers to align the dimensions of heterogeneous experts.}
    \label{fig:DMOE_architecture}
\end{figure*}

We propose a De-Correlated MoE (D-MoE), which incorporates a cross-expert de-correlation loss between expert outputs.
D-MoE follows a multi-embedding design, with each embedding table utilized by a distinct expert.
A de-correlation loss is further applied to pair-wise expert outputs to improve their specialization.
The experts are then connected through a gating mechanism, followed by an MLP to produce the final prediction, as illustrated in Fig.~\ref{fig:DMOE_architecture}.

\subsection{Embedding \& Experts}
% \paragraph{Embedding Setting: ME}
\paragraph{\textbf{Embedding}} 
% \todo{Conventional wisdom holds that we should learn one single embedding table in the MoE to share knowledge between experts.
% However, several recent works\cite{STEM2023,guo2023multi-embedding,lin2024crocodile} reveal the effectiveness of the multi-embedding paradigm, \ie, learning multiple embeddings for each feature.}
Several recent works \cite{STEM2023,guo2023multi-embedding,lin2024crocodile} demonstrate the effectiveness of the multi-embedding paradigm. Motivated by these studies, to optimally capture diverse interactions, we also adopt the multi-embedding paradigm in our D-MoE framework to facilitate the expert de-correlation, refer to Sec.~\ref{corr_analysis} for detailed evaluation.

Specifically, each expert is equipped with an exclusive embedding table to decouple feature representations explicitly.
Denote $x_i$ the raw input features of sample $i$ and expert-specific embedding tables $\{\mathbf{E}^{(m)}\}_{m=1}^M$, the embedding process is formalized as:
\begin{equation}
    \bm{e}^{(m)}_i = \mathbf{E}^{(m)}(\bm{x}_i), \quad \forall m=1,2,..., M.
\end{equation}

% Inspired by the embedding strategies applied in \cite{STEM2023,guo2023multi-embedding,lin2024crocodile}, we adopt the \textbf{multi-embedding (ME)} paradigm in our Mixture of Experts MoE framework. 
% Specifically, each expert is equipped with a private embedding table to explicitly decouple feature representations. 
% Given a raw input features of sample $i$ : $x_i$ and expert-specific embedding tables $\{\mathbf{E}^{(m)}\}_{m=1}^M$, the embedding process is formalized as:

% where $f_i$ denotes the $i$-th expert, processing its unique input embedding $e_i$.
% \mathbf{o}_i &= f_i(e_i)

\paragraph{\textbf{Experts}}
% Methods based on the Mixture of Experts (MoE) have been widely applied in multi-task~\cite{MMoE2018,PLE2020,STEM2023} and multi-domain~\cite{HMoE2020, lin2024crocodile} recommendation. 
% However, in our MoE structure, which is designed for a single task, we do not distinguish between shared experts or task-specific experts. 
% Instead, all experts operate independently and are equipped with their specific embeddings.

Given the $M$ embedding tables, we employ $M$ experts, each corresponds to one of those embedding tables.
Each expert is defined as a feature interaction model, such as a CrossNet~\cite{DCNv22021}, CIN~\cite{xDeepFM2018} or a simple DNN.
All experts can either have the same architecture, leading to a Homogeneous MoE (Homo-MoE) design, or have different architectures, leading to a Heterogeneous MoE (Hetero-MoE) design~\cite{zhang2022dhen, pan2024ads}.
Each expert processes its unique input embedding, as formalized below:
\begin{equation}
    \mathbf{o}^{(m)}_i = \text{Expert}^{(m)}(\bm{e}^{(m)}_i),
\end{equation}
where \( \text{Expert}^{(m)}(\cdot) \) denotes the \( m \)-th expert, and \( \mathbf{o}^{(m)}_i \) represents its uniquely transformed output.

% \todo{Inspired by \cite{DHEN, guo2023multi-embedding}}, we employ multiple sets of feature interaction models as experts. 
% We primarily implement homogeneous experts (\textbf{Homo-MoE}), where all experts share the same structure but have entirely independent parameters. 
% To further enhance diversity, we also explore heterogeneous experts (\textbf{Hetero-MoE}), which utilize different feature interaction models. 
% Each expert processes its unique input embedding, as formalized below:

% \subsection{Homogeneous/Heterogeneous Experts}
%  \paraholder

\subsection{De-correlation Loss}
In order to directly de-correlate the output of experts so as to capture diverse interaction on features, inspired by \cite{lin2024crocodile} and \cite{zbontar2021barlow}, we propose a cross-expert de-correlation loss (CorrLoss) to explicitly make experts de-correlated with each other.
Specifically, the loss is defined as the L2 norm of the correlation matrix between each pair of expert outputs.
Formally,
\begin{align}
\label{corrloss}
    \mathcal{L}_{Corr} = \frac{1}{d^2} \sum_{\substack{m_1, m_2 \in \{1, \ldots, M\} \\ m_1 < m_2}} \left\| \left[ \frac{\mathbf{O}^{(m_1)} - \overline{\mathbf{O}}^{(m_1)}}{\sigma(\mathbf{O}^{(m_1)})} \right]^T \left[ \frac{\mathbf{O}^{(m_2)} - \overline{\mathbf{O}}^{(m_2)}}{\sigma(\mathbf{O}^{(m_2)})} \right] \right\|_2,
\end{align}
where \(\mathbf{O}^{m_1}, \mathbf{O}^{m_2} \in \mathcal{R}^{N \times d}\) are the outputs of the \(m_1\)-th and \(m_2\)-th experts, respectively. Here, \(\mathbf{O}^{m_1} = \{\mathbf{o}^{m_1}_i\}_{i=1}^N\) aggregates the outputs of \(N\) samples from the \(m_1\)-th expert.
$\overline{\mathbf{O}}^{(m_1)} \in \mathcal{R}^{1 \times d}$ is the average value of each dimension across all samples and $\sigma(\mathbf{O}^{(p)})$ represents the standard deviation, $\|\cdot\|_2$ is the $l_2$-norm.

\subsection{Gating Mechanism}

\begin{figure}
    \centering
    \includegraphics[width=0.7\linewidth]{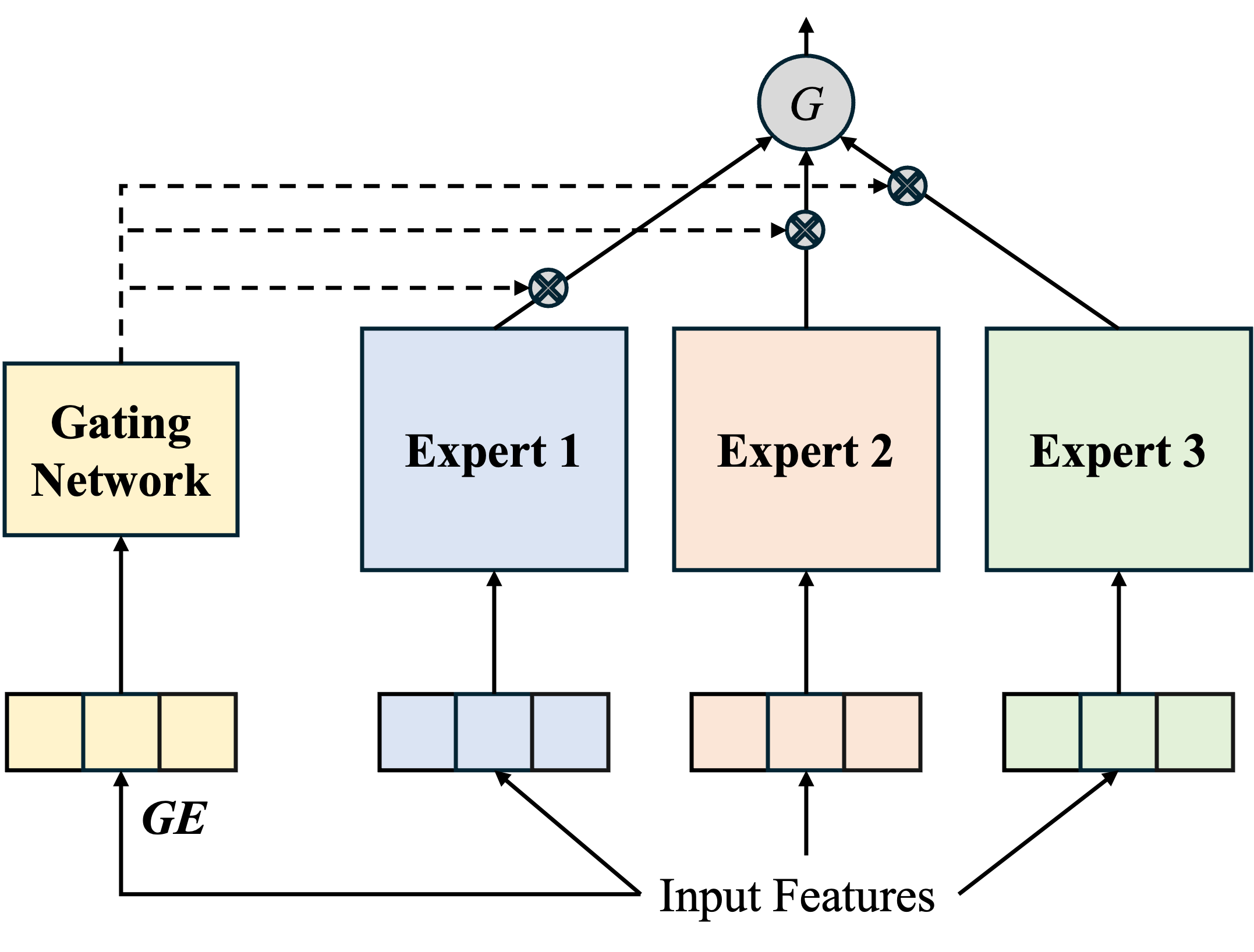}
    \vspace{-0.25cm}
    \caption{Our Gating Network mechanism}
    \label{fig:gating}
\end{figure}

The gating mechanism aims to aggregate the outputs of multiple experts and feed them to the following classifier network.
It needs to compute a gating score between the sample and each expert.
However, in our setting, each sample has multiple embeddings, and using any of them to compute the score may lead to unexpected bias towards specific experts.
To this end, \emph{we employ an expert-independent Gating Embedding ($\mathbf{GE}$) to represent each sample in the gating computation process}.

% Since our Mixture of Experts structure is based on multi-embedding, the sets of embeddings are inherently tied to the number of experts. 
% Inspired by DeepSeekMoE\cite{dai2024deepseekmoe} and the Multi-gate Mixture-of-Experts model, we propose a \textbf{Gating Network (GN)}, which is equipped with an independent set of embeddings. 
% We formulate this gating embedding as $\mathbf{GE}(x)\in \mathcal{R}^{N\times d}$, to capture the similarities between input $x$ and experts. 
% The gating embedding can be formulated as:
% \begin{equation}
%     \mathbf{E}(x)=\left[e_{0}(x), e_{1}(x), \cdots, e_{F}(x)\right]
% \end{equation}

Specifically, each sample $x_i$ is first processed through the gating embedding layer $\mathbf{GE}(x_i) \in \mathbb{R}^{d}$.
Then we compute the gating score using a Gating Network (GN), implemented as a multilayer perceptron (MLP) with ReLU activation functions.
Denoted as $W_g$, GN transforms the gating embedding $\mathbf{GE}(x_i)$ into expert-specific weights by $W_{g}(\mathbf{GE}(x_i))$. 
% To succinctly represent the GN, we define $W_g $ as the gating MLP, which transforms the gating embedding $\mathbf{GE}(x_i)$ into expert-specific weights, where $M$ represents the number of experts. 
Following the MoE paradigm in Sec.~\ref{gating}, the output of the GN is formalized as: 
\begin{align}
    g^{(m)}(\bm{x}_i) &= \text{Softmax}\left(W_{g}(\mathbf{GE}(x_i)) \right), \\
     \mathbf{h}_i &= \sum_{m=1}^M g^{(m)}(\bm{x}_i) \cdot \bm{o}_i^{(m)},
\end{align}
where $g^{(m)}(\bm{x}_i)$ denotes the gating-specific weight of the $m$-th expert, $M$ the number of experts, and the $\text{softmax}(\cdot)$ operation is applied along the second dimension to normalize expert selection probabilities. 
$h_i$ represents the aggregated output of gating.

\subsection{Tower and Loss Function}

Finally, the aggregated output $\mathbf{h}_i$ of sample \( i \), is passed through a final classifier MLP to produce the final prediction:

\begin{equation}
    \hat{y}_i = \sigma(MLP(\mathbf{h}_i))
\end{equation}

The overall loss of our \textbf{D-MoE} is formulated as follows:

\begin{equation}
   \mathcal{L}=\sum_{\mathcal{B}}\left[\frac{1}{|\mathcal{B}|} \sum_{i=1}^{|\mathcal{B}|} \mathcal{L}_{B C E}\left(\hat{y}_i, y_i\right)+\alpha \cdot \frac{1}{|\mathcal{B}|-1}\mathcal{L}_{Corr}(\{\bm{O}^{(m_i)}\}) \right]
\end{equation}

where $\mathcal{B}$ represents a batch of samples, and $\alpha$ is the weight of CorrLoss. This objective function ensures the de-correlation of every pair of experts, enhancing diversity and heterogeneity, which in turn improves overall performance.

\section{Experiments}

In this section, we provide comprehensive analyses to address the primary research questions (RQs) outlined below:
\begin{itemize}[leftmargin=*]
    \item \textbf{RQ1}: Does DMoE outperform baselines on public datasets?
    \item \textbf{RQ2}: Can CorrLoss effectively affect the expert correlation?
    \item \textbf{RQ3}: Can multi-embedding or heterogeneous expert architectures indeed achieve de-correlation? 
    \item \textbf{RQ4}: What is the optimal form of de-correlation loss?
    \item \textbf{RQ5}: Where should the de-correlation loss be applied within the DMoE architecture?
\end{itemize}

\subsection{Setup}
\subsubsection{Datasets \& evaluation protocols}
% We conduct our experiments on two widely adopted
% public CTR datasets: Avazu~\cite{avazu} and Criteo~\cite{criteo}. Dataset stastics and splits are summearized in Appendix \ref{dataset}. As for evaluation, we evaluate the recommendation performance with AUC.

We conduct our experiments on two widely recognized public datasets for CTR prediction: Avazu~\cite{avazu} and Criteo~\cite{criteo}. Detailed dataset statistics and splits are summarized in Appendix \ref{dataset}. For performance evaluation, we use metric AUC to measure predictions.

\subsubsection{Baselines} \label{baselines}
To establish a performance benchmark for comparison, we select some typical modules in CTR prediction.
Since our DMoE structure can be divided into homogeneous and heterogeneous configurations, for each structure, we select several base models and evaluate their performance.

\paragraph{\textbf{Homogeneous MoE}} 
Following~\citet{guo2023multi-embedding}, we first apply homogeneous modules in MoE structure (Homo-MoE).
We consider five typical interaction modules including NFM~\cite{NFM2017}, DeepFM~\cite{DeepFM2017}, Crossnet~\cite{DCNv22021}, DCN V2~\cite{DCNv22021}, xDeepFM~\cite{xDeepFM2018}, with their shared-embedding (SE) and multi-embedding (ME) variants as baselines. 
 
\paragraph{\textbf{Heterogeneous MoE}}
Inspired by DHEN~\cite{zhang2022dhen} and the Heterogeneous MoE with Multi-Embedding~\cite{pan2024ads}, we incorporate heterogeneous experts into the MoE structure (Hetero-MoE). 
Feasible expert modules include but are not limited to, DCN V2~\cite{DCNv22021}, CIN~\cite{xDeepFM2018}, DNN, and FM~\cite{FM2010}. Details are provided in Appendix~\ref{hetero_experts}.

We have tried different combinations of heterogeneous experts, and only report the results of the best combination.

\subsubsection{Implementation details}
For all experiments, we configure the learning rate as 0.001 and the batch size as 10000. The embedding dimensions are set to 16 for the Avazu dataset and 10 for the Criteo dataset. The MLP following the interaction model, as well as the tower network across all methods, consists of hidden units set to 500. The gate network employs an MLP with 64 hidden dimensions. The number of experts is selected from the set $\{2, 3\}$.
% We applied the two most powerful types of interaction module as fo: CIN, DCN, DNN.

\begin{table*}[ht]
\centering
\caption{AUC performance on the Avazu and Criteo datasets. Bold values highlight the best results in each model, respectively. * indicates the best AUC result for each dataset. "Single Expert" refers to the performance of a single interaction model. "SE" refers to MoE with shared-embedding and "ME" denotes as MoE equipped with multiple embedding tables.}
\renewcommand\arraystretch{1.1}
\begin{tabular}{cc|c|cccc|cccc}
\toprule
 &  & \multirow{2}{*}{\textbf{Model}} & \multicolumn{4}{c}{\textbf{Avazu}} & \multicolumn{4}{c}{\textbf{Criteo}} \\
\cmidrule(lr){4-7} \cmidrule(lr){8-11}
 &  &  & Single Expert & SE-MoE & ME-MoE & ME+Loss & Single Expert & SE-MoE & ME-MoE & ME+Loss \\
\midrule
\multirow{5}{*}{Homo-MoE} 
& & NFM & 0.789863 & 0.791739 & 0.791931 & \textbf{0.792189} & 0.804967 & 0.811121 & 0.811290 & \textbf{0.811521} \\
\cmidrule(lr){3-11}
 & & DeepFM & 0.792641 & 0.793282 & 0.793953 & \textbf{0.794002} & 0.812969 &0.813518  & 0.813322 & \textbf{0.813532} \\
\cmidrule(lr){3-11}
 & & Crossnet & 0.791229 & 0.791266 & 0.792428 & \textbf{0.792775} &0.811796 &0.811858 &0.811754 &\textbf{0.812774} \\
\cmidrule(lr){3-11}
 & & DCNv2 & 0.793308 & 0.793099 & 0.793984 & \textbf{0.794900} & 0.812559 & 0.812010 & 0.812586 & \textbf{0.812761} \\
\cmidrule(lr){3-11}
 & & xDeepFM & 0.792744 & 0.793476 & 0.793772 & \textbf{0.794692} & 0.813301 & 0.813354 &0.813706  &$\textbf{0.814010}^{*}$  \\
\midrule
\multirow{1}{*}{Hetero-MoE} & & CIN + DCN & 0.792736 & 0.793391 & 0.794553 & $\textbf{0.797236}^{*}$ &0.811511  & 0.812768 & 0.812802 & \textbf{0.812906} \\
\bottomrule
\end{tabular}
\label{tab:performance}
\end{table*}

\subsection{RQ1: Performance Evaluation}
We evaluate our Decorrelated-MoE (D-MoE) framework across various expert configurations on the Avazu~\cite{avazu} and Criteo~\cite{criteo} datasets. 
% We employed Decorrelated-MoE(D-MoE) structure in different experts setting in the Avazu\cite{avazu} and Criteo \cite{criteo} datasets. We compared the performance of metric AUC with state-of-the-art feature interaction models, shown in Table \ref{tab:performance}.
As detailed in Table \ref{tab:performance}, we first establish a baseline using a "Single Expert" structure, which represents a standalone feature interaction model. Then, incorporating multiple experts with shared embeddings (SE-MoE) further enhances model performance. Finally, the introduction of multi-embedding (ME-MoE) for each expert significantly improves results, achieving state-of-the-art performance. We compare our DMoE framework mainly against the ME-MoE baseline to demonstrate its effectiveness, with the sole variable being the implementation of \textbf{CorrLoss}. 

We can observe that applying CorrLoss consistently outperforms the state-of-the-art model (ME-MoE) across all configurations in both datasets. 
For example, on the Avazu dataset, incorporating loss with DCNv2-based ME-MoE shows a significant improvement, with the AUC increasing from 0.79398 to 0.79490, achieving a relative gain of 0.12%. 
In the Heterogeneous-MoE configuration with CIN and DCN experts, CorrLoss yields an absolute AUC improvement of 0.0027 (0.34\% relative gain). 
Such positive results can be attributed to the implementation of CorrLoss.

\subsection{RQ2: De-correlation Analysis}\label{corr_analysis}
We are curious about whether the De-correlation Loss (CorrLoss) effectively disentangles the experts as intended. 
To investigate this, we conducted a series of analysis experiments on the Avazu dataset. Our analysis progressively evaluates the effectiveness of CorrLoss, starting from an overall model-level perspective and then delving into the individual expert level. 
To determine the degree of de-correlation of experts, we propose a metric based on Pearson correlation coefficient-\textbf{Cross-expert Correlation}(CEC) as a generalized quantification.

% \begin{table}[t]
%     \centering
%     \caption{Homogeneous Experts with SE/ME with/wo Corrloss}
%     \begin{tabular}{c|cc|cc|cc}
%     \toprule
%       & \multicolumn{2}{c}{\textbf{SE}} & \multicolumn{2}{c}{\textbf{ME}} & \multicolumn{2}{c}{\textbf{ME+Loss}} \\
%     \cmidrule(lr){2-3} \cmidrule(lr){4-5} \cmidrule(lr){6-7}
%      & AUC & Corr & AUC & Corr & AUC & Corr \\
%     \hline
%     CrossNet &0.792265 &0.1690 &0.793104 &0.1040 &0.793869 &0.0024 \\
%     \hline
%     xDeepFM &0.793476 &0.1396 &0.793772 &0.1131 &0.794692 &0.0012 \\
%     \hline
%     DCNv2 &0.793099 &0.0652 &0.793984 &0.0193 &0.794900 &0.0009 \\
%     \hline
%     Hetero-MoE &0.793434 &0.0957 &0.793478 &0.0521 &0.794332 &0.0011
%     \end{tabular}
% \caption{AUC v.s. Correlation(CIN)}
% \label{table:1}
% \end{table}

\paragraph{\textbf{Definition of Metric CEC}} Consider two matrices \( X, Y \in \mathbb{R}^{N \times d} \). The Pearson Correlation Matrix \( R(X, Y) \in \mathbb{R}^{d \times d} \) is formulated as:
\begin{equation}
  R(X, Y) = \frac{1}{N-1} \left[\frac{X - \bar{X}}{\sigma(X)}\right]^T \cdot \left[\frac{Y - \bar{Y}}{\sigma(Y)}\right],
\end{equation}
where \( \bar{X} \) and \( \bar{Y} \) denote the sample means, and \( \sigma(X) \), \( \sigma(Y) \) represent the unbiased sample standard deviations of \( X \) and \( Y \), respectively.
Each element \( r_{ij} \) in \( R(X, Y) \) quantifies the Pearson correlation between the \( i \)-th dimension of \( X \) and the \( j \)-th dimension of \( Y \). The \textbf{CEC} metric between \( X \) and \( Y \) is defined as:
\begin{equation}
  \text{CEC}(X, Y) = \frac{1}{d^2} \sum_{i=1}^{d} \sum_{j=1}^{d} |r_{ij}|.
\end{equation}
Intuitively, two matrices with a low correlation metric indicate that they are de-correlated from each other.

Now we use this metric \textbf{across expert outputs} to evaluate the degree of de-correlation of experts mainly on Avazu dataset.

% \begin{table}[h]
%     \centering
%     \caption{Comparison of Metric AUC and Corr based on DMoE}
%     \begin{tabular}{c|cc|cc}
%     \toprule
%       & \multicolumn{2}{c}{\textbf{ME}} & \multicolumn{2}{c}{\textbf{ME+Loss}} \\
%     % \cmidrule(lr){2-3} \cmidrule(lr){4-5}
%      & AUC & Corr & AUC & Corr \\
%     \hline
%     CrossNet &0.793104 &0.1040 &0.793869 &0.0024 \\
%     \hline
%     xDeepFM &0.793772 &0.1131 &0.794692 &0.0012 \\
%     \hline
%     DCNv2 &0.793984 &0.0193 &0.794900 &0.0009 \\
%     \end{tabular}
%     \label{table:corr_comparison}
% \end{table}

\begin{figure}[h]
    \centering
    \includegraphics[width=0.8\linewidth]{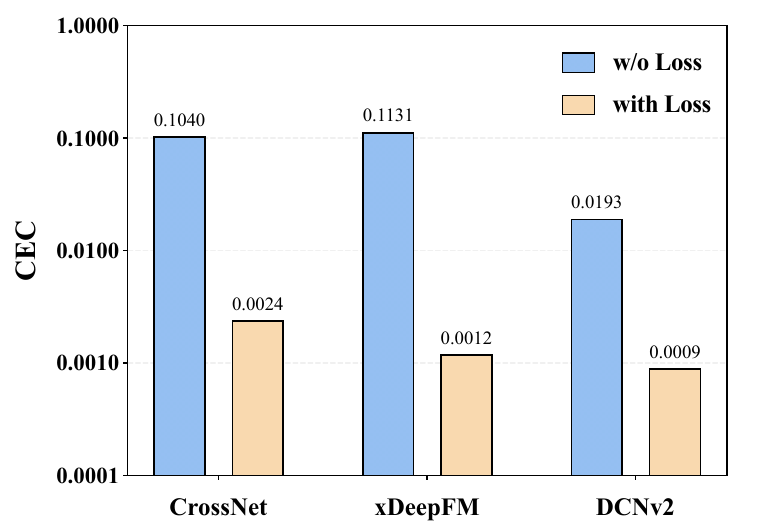}
    \vspace{-0.25cm}
    \caption{Comparison of Metric Corr, x-axis represents the feature interaction model in Homo-MoE.}
    \label{fig:corr_analysis}
\end{figure}

\textbf{Model-level Evaluation}.
First, we compute the sum of CEC metric between each pair of expert outputs in the Homo-MoE structure. 
As depicted in Fig. \ref{fig:corr_analysis}, \emph{the model integrating with CorrLoss achieves significantly lower Corr values compared to the model without CorrLoss}. 
For example, the correlation between the DCNv2 experts decreases from 0.0193 to 0.0009, representing a nearly 20-fold reduction.

\begin{figure*}[ht]
    \vspace{-0.2cm}
    \centering
    \subfloat[CEC between Homogeneous Experts\label{fig:heatmap_homo}]{%
            \includegraphics[width = 0.41\textwidth]{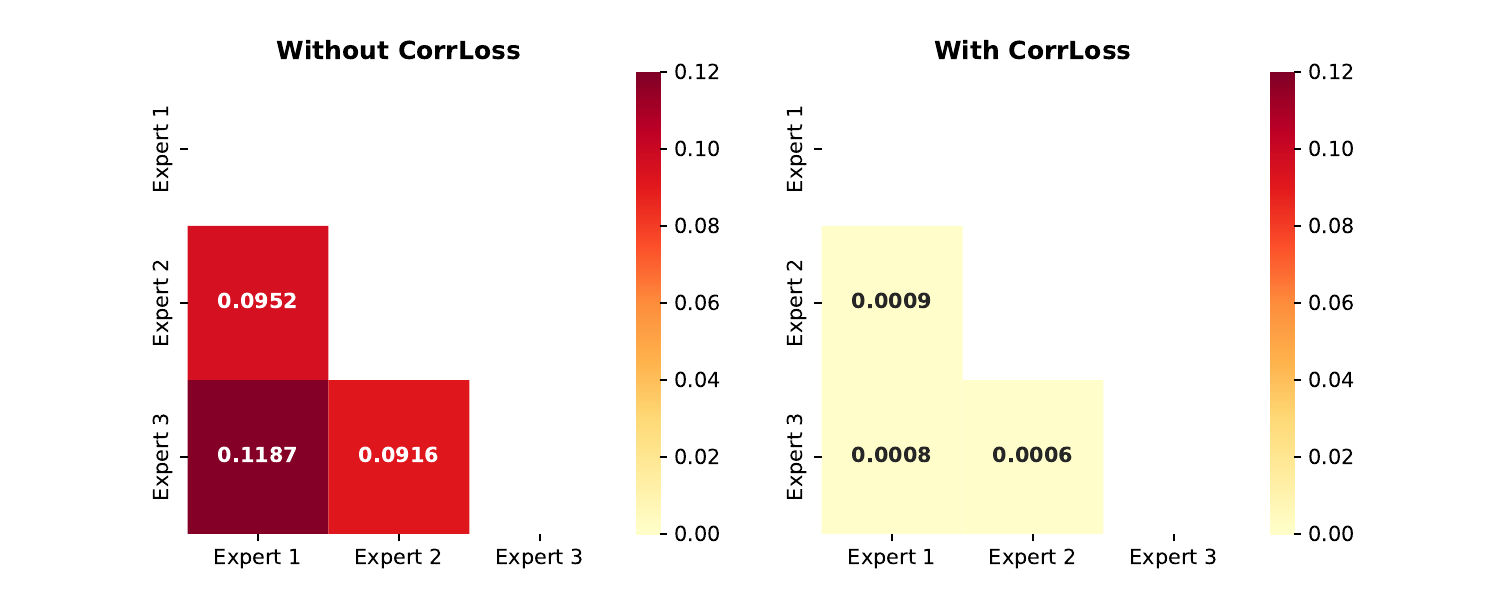}
            }
    \subfloat[CEC between Heterogeneous Experts\label{fig:heatmap_hete}]{%
            \includegraphics[width = 0.41\textwidth]{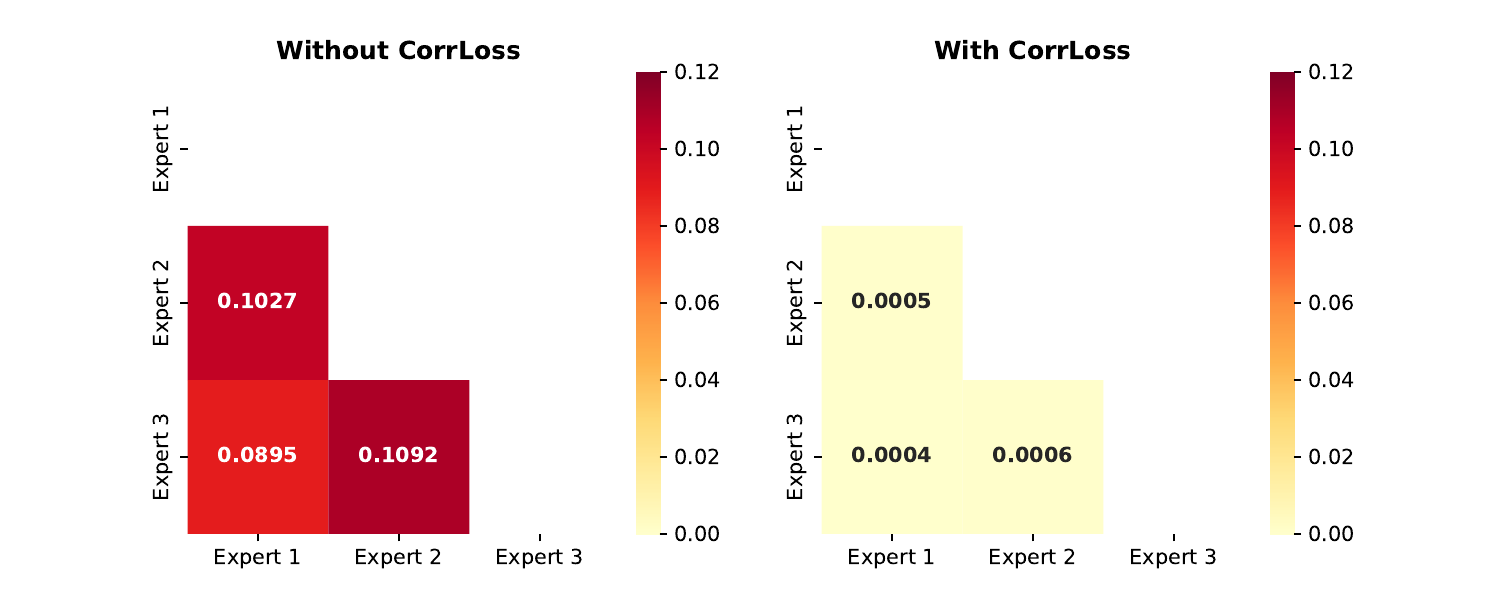}
            }
            % \hfill
    \vspace{-0.3cm}
    \caption{Heatmaps of the correlation metric (CEC) between homogeneous experts (left) and heterogeneous experts (right) on the Avazu dataset: (a) we select three xDeepFM models as homogeneous experts; (b) we choose three distinct feature interaction models as heterogeneous experts. We focus exclusively on the correlations between different experts, as illustrated by the lower triangular matrices.}
    \label{fig:heatmaps_comparison}
\end{figure*}

\textbf{Expert-level Evaluation}.
Secondly, we separately calculate the Corr between each pair of distinct experts, using three experts as representatives. 
As shown in Figure \ref{fig:heatmaps_comparison}, our method with CorrLoss substantially reduces the CEC metrics across each pair of experts in both Homo-MoE and Hetero-MoE structures. 
For instance, the correlation between Expert-1 and Expert-2 in Homo-MoE decreases approximately 1000-fold, from 0.0952 to 0.0009.

Building on our analysis of expert de-correlation, the proposed CorrLoss effectively reduces dependencies between expert outputs, as quantified by the CEC metric. 
When implemented with the de-correlation principle, CorrLoss achieves notable performance gains: 0.13\% AUC improvement in the Homo-MoE architecture and 0.22\% in the Hetero-MoE framework. 
These results demonstrate that CorrLoss successfully promotes greater differentiation among experts. while simultaneously enhancing model efficacy.

\subsection{RQ3: De-correlation Analysis on other methods} \label{method_analysis}
In this section, we investigate alternative approaches to de-correlated experts besides employing the de-correlation loss.
As discussed in Sec.~\ref{2.3}, multi-embedding(ME) aims to achieve de-correlation at the input level, while heterogeneous MoE(Hetero-MoE) leverages diverse experts to capture distinct feature interaction patterns. To validate whether these approaches effectively reduce correlations, we employ the CEC metric in Section \ref{corr_analysis} across the output of experts for quantitative evaluation.

We design our experiments using a module comprising two homogeneous experts with shared embedding(SE). In the multi-embedding configuration, we retain the same expert architecture but introduce a new set of embedding tables to de-correlate at the input level. For the heterogeneous experts setting, we replace one of the homogeneous experts with a heterogeneous expert, thereby diversifying the feature interaction patterns. 
We conduct experiments on two interaction models: DCN and CIN, as discussed in Sec.~\ref{baselines}. The heterogeneous framework integrates both models.
% \begin{figure}[h]
%     \centering
%     \begin{subfigure}[t]{0.23\textwidth} 
%         \includegraphics[width=\textwidth]{figure/cin.png}
%         \caption{CIN}
%         \label{fig:cin}
%     \end{subfigure}
%     % \hspace{0.05\textwidth} % Adds horizontal space between the subfigures
%     \begin{subfigure}[t]{0.23\textwidth} 
%         \includegraphics[width=\textwidth]{figure/dcn.png}
%         \caption{DCN}
%         \label{fig:dcn}
%     \end{subfigure}
%     \vspace{-0.25cm}
%     \caption{Comparison of AUC and the CEC metric for two decorrelation methods—multi-embedding (ME) and heterogeneous experts (Hetero-MoE) based on CIN (left) and DCN (right). Lower CEC values indicate stronger decorrelation between experts.}
%     \label{fig:ablation}
% \end{figure}

\begin{figure}[h]
    \centering
    \subfloat[CIN\label{fig:cin}]{%
            \includegraphics[width = 0.24\textwidth]{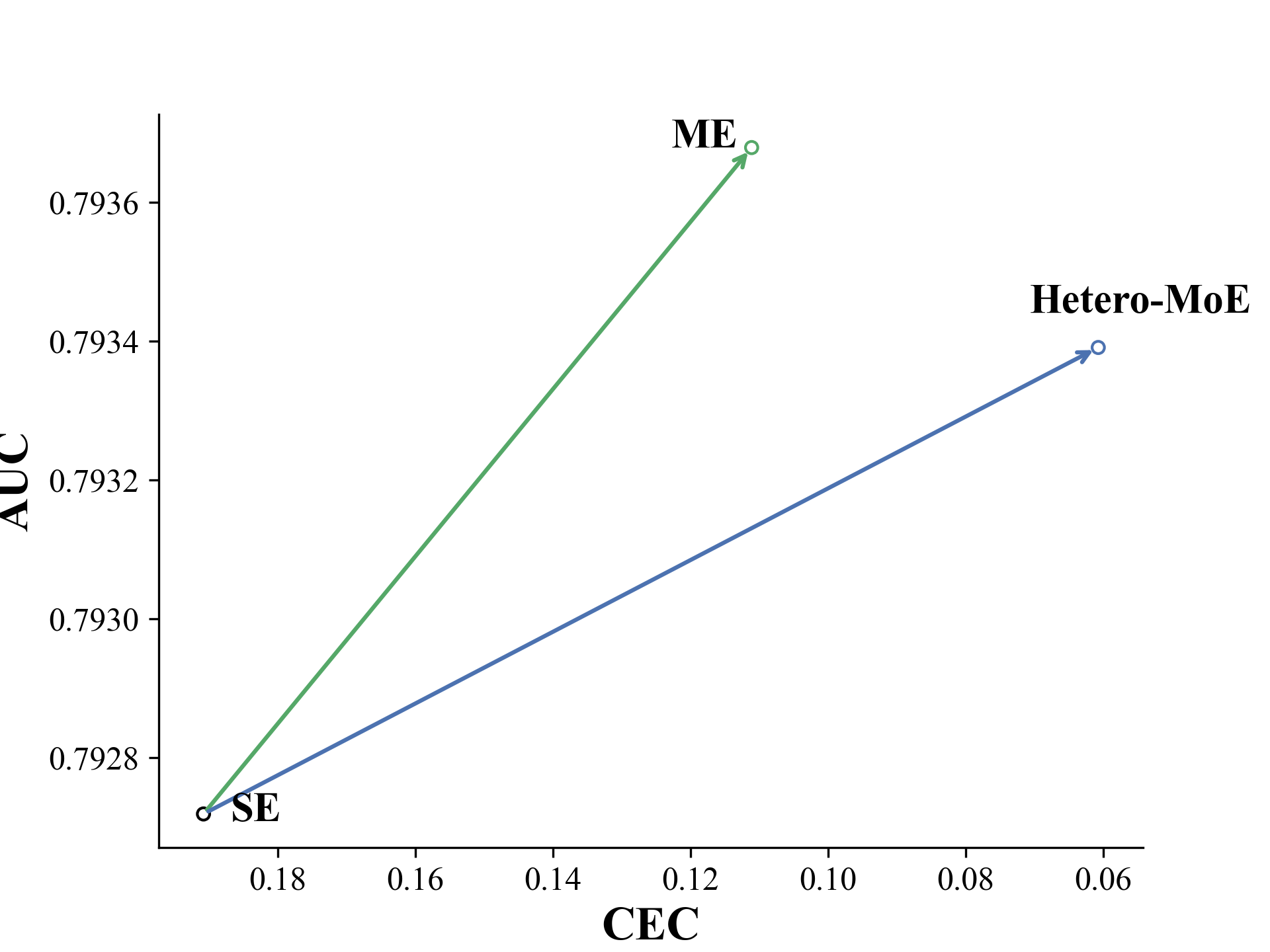}
            }
    \subfloat[DCN\label{fig:dcn}]{%
            \includegraphics[width = 0.24\textwidth]{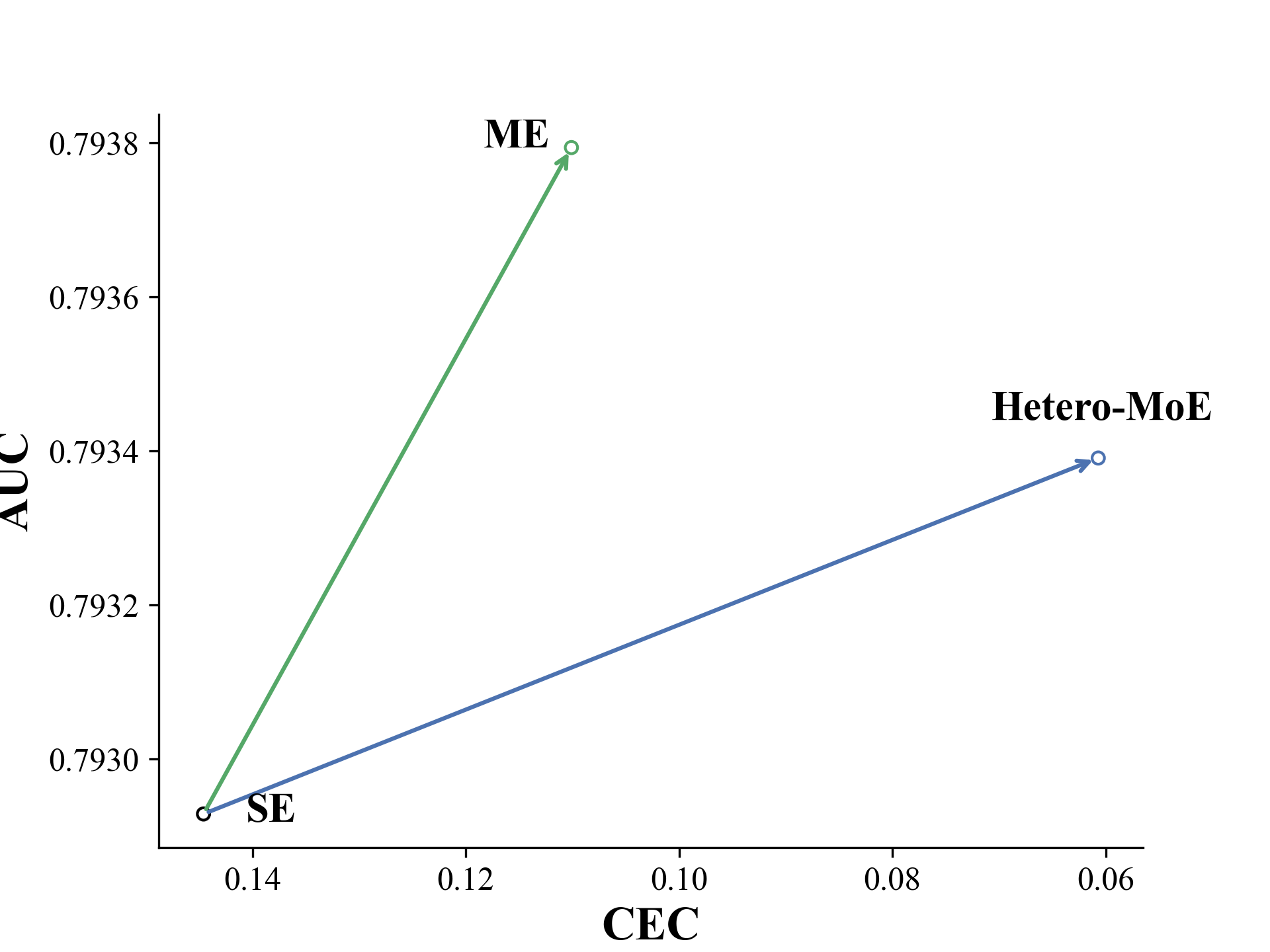}
            }
            % \hfill
    \vspace{-0.25cm}
    \caption{Comparison of AUC and the CEC metric for two decorrelation methods—multi-embedding (ME) and heterogeneous experts (Hetero-MoE) based on CIN (left) and DCN (right). Lower CEC values indicate stronger decorrelation between experts. }
    \label{fig:ablation}
\end{figure}
We visualize model performance(AUC) alongside the correlation metric (CEC) in Fig.~\ref{fig:ablation}, where higher AUC represents better performance while lower Corr indicate a higher de-correlation. 

\paragraph{\textbf{ME}} As shown in Fig. \ref{fig:cin} (green line), applying novel embeddings to the CIN-based MoE reduces cross-expert correlation from 0.1908 to 0.1111 while marginally improving AUC by approximately 0.12\%. This trend is replicated in the DCN-based MoE, supporting our hypothesis that multi-embedding strategies effectively de-correlate experts while enhancing input diversity.

\paragraph{\textbf{Hetero-MoE}} As indicated by the blue line in Fig. \ref{fig:cin} and \ref{fig:dcn}, replacing a homogeneous expert with a heterogeneous one results in a lower CEC metric and improved model performance. For example, replacing a DCN expert with a CIN expert in a DCN-based MoE decreases the cross-expert correlation from 0.1446 to 0.0607, indicating that increasing expert heterogeneity significantly lowers cross-expert correlations.

As previously mentioned in Sec.\ref{2.3}, both the multi-embedding (ME) and heterogeneous MoE (Hetero-MoE) approaches inherently achieve de-correlation, as demonstrated by their effectiveness in reducing cross-expert correlations.

\subsection{De-correlation Principle}
Based on our experimental findings, we evaluate three decorrelation strategies: multi-embedding (ME), heterogeneous experts, and decorrelation loss (CorrLoss). Each approach effectively reduces correlations across experts while improving AUC. We progressively integrate these strategies.
We implement CIN and DCN as the base interaction models for Homo-MoE, and then Hetero-MoE combines both modules. We visualized the progression on the Avazu dataset in Fig.~\ref{fig:intro}. 
It is worth noting that the degree of decorrelation is inversely proportional to the CEC metric.

We offer more details regarding this figure and make the following observations: 
1) Starting from a shared-embedding MoE with homogeneous experts, employing the ME paradigm results in an AUC improvement of approximately \(0.12\%\) and \(0.11\%\) on CIN and DCN, respectively. Meanwhile, the CEC reduces from \(0.169\) to \(0.104\) in CIN and from \(0.191\) to \(0.111\) in DCN.
2) Further employing the de-correlation loss on Homo-MoE can progressively improve the AUC by \(0.10\%\) and \(0.17\%\), respectively, and significantly reduce the correlation, decreasing by \(1.5e^{-2}\) on both models.
3) Same with illustration in Sec.~\ref{method_analysis}, substituting with a heterogeneous expert results in better performance and lower CEC. Furthermore, based on a SE-MoE with heterogeneous experts, employing the ME paradigm also exhibits an AUC improvement of \(0.12\%\) and a decline of CEC from \(0.082\) to \(0.074\), indicating that employing the ME paradigm in Hetero-MoE is also effective.
4) Based on a Hetero-MoE, the integration of CorrLoss further boosts AUC from \(0.7941\) to \(0.7971\) and reduces CEC by a factor of \(10^{-2}\), demonstrating the effectiveness of the de-correlation loss.

These results demonstrate the compatibility of de-correlation strategies. Also support the principle that progessively de-correlation across experts enhances model performance.

\subsection{RQ4: Formulation of De-correlated Loss}
There are several methods to de-correlate representations, typically by calculating covariance or Pearson's correlation coefficient.
We referenced the loss functions from \cite{lin2024crocodile} and Barlow Twins\cite{zbontar2021barlow}, simplified them into two types of losses: covariance-based loss and correlation-based loss.
The covariance-based loss can be further categorized into $Cov(L_1)$ and $Cov(L_2)$ based on the $L_1$ and $L_2$ norms.
Specifically, we calculate the loss on expert $p$ and $q$ as follows.
\begin{align}
\label{covloss_l1}
\mathcal{L}_{Cov(L_1)} = \frac{1}{d^2}||[\mathbf{O}^{(p)}-\overline{\mathbf{O}}^{(p)}]^T[\mathbf{O}^{(q)}-\overline{\mathbf{O}}^{(q)}]||_1,
\end{align}
\begin{align}
\label{covloss_l2}
\mathcal{L}_{Cov(L_2)} = \frac{1}{d^2}||[\mathbf{O}^{(p)}-\overline{\mathbf{O}}^{(p)}]^T[\mathbf{O}^{(q)}-\overline{\mathbf{O}}^{(q)}]||_2,
\end{align}
\begin{align}
\label{corrloss}
\mathcal{L}_{Corr} = \frac{1}{d^2}\left\| \left[ \frac{\mathbf{O}^{(p)} - \overline{\mathbf{O}}^{(p)}}{\sigma(\mathbf{O}^{(p)})} \right]^T \left[ \frac{\mathbf{O}^{(q)} - \overline{\mathbf{O}}^{(q)}}{\sigma(\mathbf{O}^{(q)})} \right] \right\|_2,
\end{align}

\begin{figure}[h]
    \centering
    \includegraphics[width=0.6\linewidth]{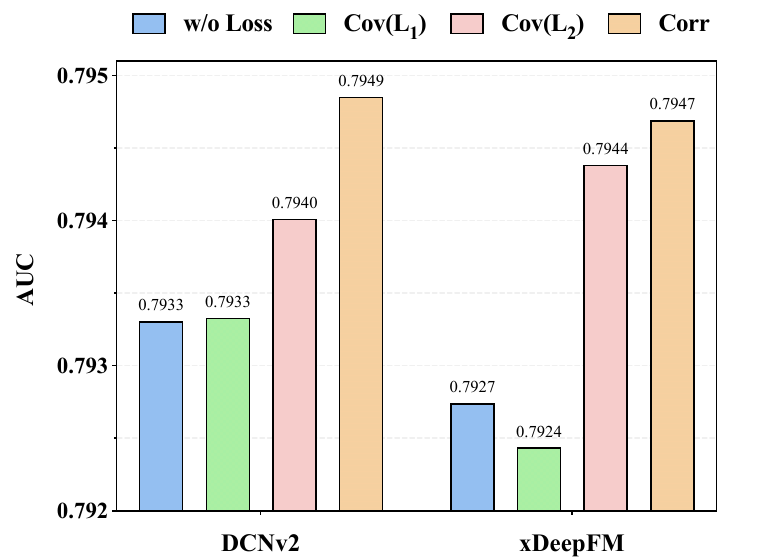}
    \vspace{-0.25cm}
    \caption{Formulation of loss, x-axis represents the feature interaction model in Homo-MoE structure.}
    \label{fig:loss_form}
\end{figure}

To comprehensively evaluate the effectiveness of different formulations of the loss function, we conduct experiments on two model architectures: ME-DCNv2 and ME-xDeepFM, each equipped with two experts. 
The experimental results are summarized in Fig. ~\ref{fig:loss_form}.

% \begin{table}[h]
%     \centering
%     \caption{Formulation of loss(use bars figure to illustrate)}
%     \begin{tabular}{c|c|c}
%     \toprule
%      & DCNv2 &xDeepFM  \\
%     \midrule
%     w/o Loss  &0.793308 &0.792744 \\
%     \midrule
%      Cov($L_1$)  &0.793333 &0.792439 \\
%      \midrule
%      Cov($L_2$)  &0.794013 &0.794385 \\
%      \midrule
%      Corr      &\textbf{0.794853} &\textbf{0.794692}  \\
%     \bottomrule
%     \end{tabular}
%     \label{tab:loss_form}
% \end{table}

Compared with covariance-based loss, our correlation-based loss demonstrates superior performance in both configurations. 
For instance, applying CorrLoss improves AUC by approximately 0.20\% compared to the baseline without loss, whereas CovLoss achieves only a 0.08\% improvement in the DCNv2 configuration. Furthermore, in the xDeepFM framework, the covariance-based loss under the $L_1$ norm adversely affects model performance.

% In summary, our proposed correlation-based loss $\mathcal{L}_{Corr}$ is empirically validated to be effective in enhancing model performance. By explicitly minimizing the correlation between expert outputs, our approach ensures that each expert contributes specialized and complementary information, leading to accurate predictions.

% \subsection{RQ5: Location of Applying De-correlated Loss}
\subsection{RQ5: Location of Applying CorrLoss}
After defining the formulation of the De-correlation Loss, we explore its optimal application location within the model architecture. Specifically, we investigate three potential locations to apply the loss:
1) \textbf{Input}: We apply the loss to the input of each expert, aiming to de-correlate the inputs at the initial stage.
2) \textbf{Intermediate layers}: We apply the loss to intermediate layers of the model, such as the $l$-th layer of the Crossnet, de-correlating the internal representations of the experts. 
3) \textbf{Output}: We apply the loss to the output of each expert, directly de-correlating the outputs of different experts.

We conduct experiments based on two Homo-MoE structures: ME-Crossnet and ME-DCNv2, each equipped with two experts. For Loss on intermediate layers, we apply on every cross layers of 2 experts.
The experimental results can be referenced in Tab. ~\ref{tab:loss_position}.

\begin{table}[h]
    \centering
    \caption{AUC performance on the Avazu dataset while applying CorrLoss at different locations.}
    \begin{tabular}{c|c|c}
    \toprule
     & CrossNet &DCNv2  \\
    \midrule
    w/o Loss  &0.791229 &0.793308 \\
    \midrule
     Input  &0.792439 &0.794463 \\
     \midrule
     Intermediate layers  &0.792688 &0.794438 \\
     \midrule
     \rowcolor[HTML]{CCFFCC}
     Output &\textbf{0.792775} &\textbf{0.794900} \\
    \bottomrule
    \end{tabular}
    \label{tab:loss_position}
\end{table}

Our experimental results demonstrate that applying the \textbf{De-correlation Loss} at distinct locations: \textbf{input}, \textbf{intermediate layers}, and \textbf{output} consistently enhances model performance compared to the base model (without the loss). Specifically:
For the ME-CrossNet model, applying the loss at the input layer increases the AUC from 0.79123 to 0.79244. 
Introducing the loss at the intermediate layers further elevates the AUC to 0.79269, while the output layer achieves the highest AUC of 0.79278, corresponding to an overall improvement of 0.2\% .
ME-DCNv2 exhibits a similar trend, with applying loss at experts output yielding the highest AUC of 0.7949.
In summary, while all three approaches surpass the baseline, \emph{applying the loss at the experts’ output layer achieves best performance}.
% , emphasizing the necessity of diversifying final expert representations. 
% This finding highlights the optimal strategy for de-correlation, which is exactly employed in our DMoE framework.

\section{Online A/B Testing}
We deployed the proposed De-correlated MoE in one of the world's largest advertising platforms. First, almost all pCTR models in our production platform employ Heterogeneous Experts with Multi-Embedding architecture~\cite{guo2023multi-embedding, STEM2023, pan2024ads}. In particular, these heterogeneous experts use different feature interaction functions, \textit{e.g.}, IPNN~\cite{PNN2016}, DCNv2~\cite{DCNv22021}, FlatDNN, Wukong~\cite{zhang2024wukong}, and GwPFM~\cite{pan2024ads}, with several embedding tables. For example, the Moments pCTR model consists of a GwPFM, IPNN, and FlatDNN, along with two embedding tables. GwPFM and FlatDNN share the first table, while IPNN uses the second one.  Meanwhile, there are at most $8$ fully connected layers in each expert while the hidden dimension of each layer is as large as $2028$. Further, the embedding size ranges from $8$ to $64$ for more than five hundred user-, ad-, and context-side features to capture the interests of diverse users.

We deployed the proposed De-correlated MoE in our production platform by introducing a de-correlation loss for each pair of experts to explicitly de-correlate these experts. During the one-week 20\% A/B testing, promising performances are witnessed, achieving
significant GMV lift ranging from 1.02\% to 2.68\%, as well as CTR lift ranging from 0.95\% to 2.46\%, on several vital scenarios, including Channel pCTR, Moments pCTR, Content and Platform pCTR, and DSP pCTR. These improvements were statistically significant according to t-tests. Finally, the proposed De-correlated MoE has been successfully deployed as the production model in the above-mentioned scenarios, leading to a revenue lift by hundreds of millions of dollars per year.

\section{Related works}
\subsection{Feature interaction models}
Explicit interaction models, particularly the 2nd-order ones, have well-defined definitions and close-form formulations. 
This line of models originated from classic Matrix Factorization (MF)~\cite{MF2009}, followed by FM~\cite{FM2010}, FFM~\cite{FFM2016}, FwFM~\cite{FwFM2018}, and FmFM~\cite{FmFM2021}. 
In contrast, models like xDeepFM~\cite{xDeepFM2018} and DCNv2~\cite{DCNv22021} employ a parameter or weight matrix to model higher-order interactions, going beyond the 2nd-order interactions captured by the aforementioned models.
\cite{kang2024unified_framework} summarize existing feature interaction functions from the simple dot-product~\cite{FM2010} to the most complicated projected product~\cite{FibiNet2019, FmFM2021, DCNv22021}. 
It shows that the more complex the interaction functions, the more robust the learned embeddings.

\subsection{MoE in recommender systems}
Initially introduced by~\citet{jacobs1991adaptive}, Mixture-of-Experts (MoE) have attracted wide attention from academic and industrial areas~\cite{jacobs1991adaptive, eigen2013learning, shazeer2017outrageously, dai2024deepseekmoe}.
MoE architecture has also been widely adopted in recommendation systems, especially in in multi-task or multi-domain(or scenario) recommendation~\cite{moe, moe_survey, MMoE2018, PLE2020, STEM2023, lin2024crocodile, llm_survey, me_dhen}.
Specifically, when learning several tasks or domains simultaneously, there are inherent conflicts between them, leading to negative transfer or seesaw phenomenon.
Consequently, many works proposed to employ several towers in the top, one for each task (group) or domain (group), and multiple experts in the bottom.
Then a MoE like routing between there towers and experts are adopted, often referred to as the Multi-gate Mixture-of-Experts architecture.

For example, MMoE~\cite{MMoE2018, HMoE2020} proposed to learn several symmetric experts to capture the common knowledge between tasks/domains.
PLE~\cite{PLE2020} proposed to learn task-specific experts to learn the distinctiveness of each task.
STEM~\cite{STEM2023} proposed to learn task-specific embedding tables and experts to further decouple the effect of tasks.
Crocodile~\cite{lin2024crocodile} proposed to learn several symmetric yet de-correlated experts to learn the common knowledge as well as the distinctiveness of domains.
Besides, there are also several recent work to employ MoE for single-task CTR prediction task.
~\citet{zhang2022dhen} proposed a Deep and Hierarchical Ensemble Network to learn several heterogeneous experts.
~\citet{pan2024ads} proposed a Heterogeneous Mixture-of-Experts with Multi-Embedding.

\section{Conclusion}
In this paper, we have systematically studied the MoE de-correlation problem for recommender systems.
We have derived a novel metric termed \emph{CEC}, which exhibits a strong connection with downstream recommendation performance and can be used to quantitatively investigate the de-correlation degree among MoE experts.
We proposed a De-Correlated MoE (D-MoE), which incorporates a Cross-Expert De-Correlation loss between the outputs of the experts.
We have conducted comprehensive evaluations on two popular public datasets to verify the superiority of D-MoE.
Furthermore, we find a general principle: employing various de-correlation approaches such as the Multi-Embedding paradigm and the De-Correlation loss can progressively de-correlate the expert and enhance model performance.
% In future work, we will explore DMoE with sparse routing mechanism.

\clearpage
\bibliographystyle{ACM-Reference-Format}
\bibliography{8.reference}


\begin{thebibliography}{45}

%%% the Web address.

\ifx \showCODEN    \undefined \def \showCODEN     #1{\unskip}     \fi
\ifx \showISBNx    \undefined \def \showISBNx     #1{\unskip}     \fi
\ifx \showISBNxiii \undefined \def \showISBNxiii  #1{\unskip}     \fi
\ifx \showISSN     \undefined \def \showISSN      #1{\unskip}     \fi
\ifx \showLCCN     \undefined \def \showLCCN      #1{\unskip}     \fi
\ifx \shownote     \undefined \def \shownote      #1{#1}          \fi
\ifx \showarticletitle \undefined \def \showarticletitle #1{#1}   \fi
\ifx \showURL      \undefined \def \showURL       {\relax}        \fi
% invisible to TeX
\providecommand\bibfield[2]{#2}
\providecommand\bibinfo[2]{#2}
\providecommand\natexlab[1]{#1}
\providecommand\showeprint[2][]{arXiv:#2}

\bibitem[ava(2014)]%
        {avazu}
 \bibinfo{year}{2014}\natexlab{}.
\newblock \bibinfo{title}{Avazu Dataset}.
\newblock \bibinfo{howpublished}{\url{https://www.kaggle.com/competitions/avazu-ctr-prediction/data}}.
\newblock


\bibitem[cri(2014)]%
        {criteo}
 \bibinfo{year}{2014}\natexlab{}.
\newblock \bibinfo{title}{Criteo Dataset}.
\newblock \bibinfo{howpublished}{\url{https://www.kaggle.com/c/criteo-display-ad-challenge/data}}.
\newblock


\bibitem[Cai et~al\mbox{.}(2024)]%
        {moe_survey}
\bibfield{author}{\bibinfo{person}{Weilin Cai}, \bibinfo{person}{Juyong Jiang}, \bibinfo{person}{Fan Wang}, \bibinfo{person}{Jing Tang}, \bibinfo{person}{Sunghun Kim}, {and} \bibinfo{person}{Jiayi Huang}.} \bibinfo{year}{2024}\natexlab{}.
\newblock \showarticletitle{A survey on mixture of experts}.
\newblock \bibinfo{journal}{\emph{arXiv preprint arXiv:2407.06204}} (\bibinfo{year}{2024}).
\newblock


\bibitem[Caruana(1997)]%
        {SharedBottom1997}
\bibfield{author}{\bibinfo{person}{Rich Caruana}.} \bibinfo{year}{1997}\natexlab{}.
\newblock \showarticletitle{Multitask learning}.
\newblock \bibinfo{journal}{\emph{Machine learning}} \bibinfo{volume}{28}, \bibinfo{number}{1} (\bibinfo{year}{1997}), \bibinfo{pages}{41--75}.
\newblock


\bibitem[Dai et~al\mbox{.}(2024)]%
        {dai2024deepseekmoe}
\bibfield{author}{\bibinfo{person}{Damai Dai}, \bibinfo{person}{Chengqi Deng}, \bibinfo{person}{Chenggang Zhao}, \bibinfo{person}{RX Xu}, \bibinfo{person}{Huazuo Gao}, \bibinfo{person}{Deli Chen}, \bibinfo{person}{Jiashi Li}, \bibinfo{person}{Wangding Zeng}, \bibinfo{person}{Xingkai Yu}, \bibinfo{person}{Y Wu}, {et~al\mbox{.}}} \bibinfo{year}{2024}\natexlab{}.
\newblock \showarticletitle{Deepseekmoe: Towards ultimate expert specialization in mixture-of-experts language models}.
\newblock \bibinfo{journal}{\emph{arXiv preprint arXiv:2401.06066}} (\bibinfo{year}{2024}).
\newblock


\bibitem[Eigen et~al\mbox{.}(2013)]%
        {eigen2013learning}
\bibfield{author}{\bibinfo{person}{David Eigen}, \bibinfo{person}{Marc'Aurelio Ranzato}, {and} \bibinfo{person}{Ilya Sutskever}.} \bibinfo{year}{2013}\natexlab{}.
\newblock \showarticletitle{Learning factored representations in a deep mixture of experts}.
\newblock \bibinfo{journal}{\emph{arXiv preprint arXiv:1312.4314}} (\bibinfo{year}{2013}).
\newblock


\bibitem[Guo et~al\mbox{.}(2017)]%
        {DeepFM2017}
\bibfield{author}{\bibinfo{person}{Huifeng Guo}, \bibinfo{person}{Ruiming Tang}, \bibinfo{person}{Yunming Ye}, \bibinfo{person}{Zhenguo Li}, {and} \bibinfo{person}{Xiuqiang He}.} \bibinfo{year}{2017}\natexlab{}.
\newblock \showarticletitle{DeepFM: a factorization-machine based neural network for CTR prediction}.
\newblock \bibinfo{journal}{\emph{arXiv preprint arXiv:1703.04247}} (\bibinfo{year}{2017}).
\newblock


\bibitem[Guo et~al\mbox{.}(2024)]%
        {guo2023multi-embedding}
\bibfield{author}{\bibinfo{person}{Xingzhuo Guo}, \bibinfo{person}{Junwei Pan}, \bibinfo{person}{Ximei Wang}, \bibinfo{person}{Baixu Chen}, \bibinfo{person}{Jie Jiang}, {and} \bibinfo{person}{Mingsheng Long}.} \bibinfo{year}{2024}\natexlab{}.
\newblock \showarticletitle{On the Embedding Collapse when Scaling up Recommendation Models}.
\newblock \bibinfo{journal}{\emph{International Conference on Machine Learning (ICML)}} (\bibinfo{year}{2024}).
\newblock


\bibitem[He and Chua(2017)]%
        {NFM2017}
\bibfield{author}{\bibinfo{person}{Xiangnan He} {and} \bibinfo{person}{Tat-Seng Chua}.} \bibinfo{year}{2017}\natexlab{}.
\newblock \showarticletitle{Neural factorization machines for sparse predictive analytics}. In \bibinfo{booktitle}{\emph{Proceedings of the 40th International ACM SIGIR conference on Research and Development in Information Retrieval}}. \bibinfo{pages}{355--364}.
\newblock


\bibitem[Huang et~al\mbox{.}(2019)]%
        {FibiNet2019}
\bibfield{author}{\bibinfo{person}{Tongwen Huang}, \bibinfo{person}{Zhiqi Zhang}, {and} \bibinfo{person}{Junlin Zhang}.} \bibinfo{year}{2019}\natexlab{}.
\newblock \showarticletitle{FiBiNET: combining feature importance and bilinear feature interaction for click-through rate prediction}. In \bibinfo{booktitle}{\emph{Proceedings of the 13th ACM Conference on Recommender Systems}}. \bibinfo{pages}{169--177}.
\newblock


\bibitem[Jacobs et~al\mbox{.}(1991)]%
        {jacobs1991adaptive}
\bibfield{author}{\bibinfo{person}{Robert~A Jacobs}, \bibinfo{person}{Michael~I Jordan}, \bibinfo{person}{Steven~J Nowlan}, {and} \bibinfo{person}{Geoffrey~E Hinton}.} \bibinfo{year}{1991}\natexlab{}.
\newblock \showarticletitle{Adaptive mixtures of local experts}.
\newblock \bibinfo{journal}{\emph{Neural computation}} \bibinfo{volume}{3}, \bibinfo{number}{1} (\bibinfo{year}{1991}), \bibinfo{pages}{79--87}.
\newblock


\bibitem[Juan et~al\mbox{.}(2016)]%
        {FFM2016}
\bibfield{author}{\bibinfo{person}{Yuchin Juan}, \bibinfo{person}{Yong Zhuang}, \bibinfo{person}{Wei-Sheng Chin}, {and} \bibinfo{person}{Chih-Jen Lin}.} \bibinfo{year}{2016}\natexlab{}.
\newblock \showarticletitle{Field-aware factorization machines for CTR prediction}. In \bibinfo{booktitle}{\emph{Proceedings of the 10th ACM Conference on Recommender Systems (RecSys)}}. \bibinfo{pages}{43--50}.
\newblock


\bibitem[Kang et~al\mbox{.}(2024)]%
        {kang2024unified_framework}
\bibfield{author}{\bibinfo{person}{Yu Kang}, \bibinfo{person}{Junwei Pan}, \bibinfo{person}{Jipeng Jin}, \bibinfo{person}{Shudong Huang}, \bibinfo{person}{Xiaofeng Gao}, {and} \bibinfo{person}{Lei Xiao}.} \bibinfo{year}{2024}\natexlab{}.
\newblock \showarticletitle{A Unified Framework of Feature Interaction Models for Recommendation}.
\newblock \bibinfo{journal}{\emph{arxiv}} (\bibinfo{year}{2024}).
\newblock


\bibitem[Koren et~al\mbox{.}(2009)]%
        {MF2009}
\bibfield{author}{\bibinfo{person}{Yehuda Koren}, \bibinfo{person}{Robert Bell}, {and} \bibinfo{person}{Chris Volinsky}.} \bibinfo{year}{2009}\natexlab{}.
\newblock \showarticletitle{Matrix factorization techniques for recommender systems}.
\newblock \bibinfo{journal}{\emph{Computer}} \bibinfo{volume}{42}, \bibinfo{number}{8} (\bibinfo{year}{2009}), \bibinfo{pages}{30--37}.
\newblock


\bibitem[Li et~al\mbox{.}(2024)]%
        {dcnv3}
\bibfield{author}{\bibinfo{person}{Honghao Li}, \bibinfo{person}{Yiwen Zhang}, \bibinfo{person}{Yi Zhang}, \bibinfo{person}{Hanwei Li}, \bibinfo{person}{Lei Sang}, {and} \bibinfo{person}{Jieming Zhu}.} \bibinfo{year}{2024}\natexlab{}.
\newblock \showarticletitle{DCNv3: Towards Next Generation Deep Cross Network for CTR Prediction}.
\newblock \bibinfo{journal}{\emph{arXiv preprint arXiv:2407.13349}} (\bibinfo{year}{2024}).
\newblock


\bibitem[Li et~al\mbox{.}(2020)]%
        {HMoE2020}
\bibfield{author}{\bibinfo{person}{Pengcheng Li}, \bibinfo{person}{Runze Li}, \bibinfo{person}{Qing Da}, \bibinfo{person}{An-Xiang Zeng}, {and} \bibinfo{person}{Lijun Zhang}.} \bibinfo{year}{2020}\natexlab{}.
\newblock \showarticletitle{Improving multi-scenario learning to rank in e-commerce by exploiting task relationships in the label space}. In \bibinfo{booktitle}{\emph{Proceedings of the 29th ACM International Conference on Information \& Knowledge Management}}. \bibinfo{pages}{2605--2612}.
\newblock


\bibitem[Lian et~al\mbox{.}(2018)]%
        {xDeepFM2018}
\bibfield{author}{\bibinfo{person}{Jianxun Lian}, \bibinfo{person}{Xiaohuan Zhou}, \bibinfo{person}{Fuzheng Zhang}, \bibinfo{person}{Zhongxia Chen}, \bibinfo{person}{Xing Xie}, {and} \bibinfo{person}{Guangzhong Sun}.} \bibinfo{year}{2018}\natexlab{}.
\newblock \showarticletitle{xdeepfm: Combining explicit and implicit feature interactions for recommender systems}. In \bibinfo{booktitle}{\emph{Proceedings of the 24th ACM SIGKDD International Conference on Knowledge Discovery and Data Mining (SIGKDD)}}. \bibinfo{pages}{1754--1763}.
\newblock


\bibitem[Lin et~al\mbox{.}(2024)]%
        {lin2024crocodile}
\bibfield{author}{\bibinfo{person}{Zhutian Lin}, \bibinfo{person}{Junwei Pan}, \bibinfo{person}{Haibin Yu}, \bibinfo{person}{Xi Xiao}, \bibinfo{person}{Ximei Wang}, \bibinfo{person}{Zhixiang Feng}, \bibinfo{person}{Shifeng Wen}, \bibinfo{person}{Shudong Huang}, \bibinfo{person}{Lei Xiao}, {and} \bibinfo{person}{Jie Jiang}.} \bibinfo{year}{2024}\natexlab{}.
\newblock \showarticletitle{Crocodile: Cross Experts Covariance for Disentangled Learning in Multi-Domain Recommendation}.
\newblock \bibinfo{journal}{\emph{arXiv preprint arXiv:2405.12706}} (\bibinfo{year}{2024}).
\newblock


\bibitem[Liu et~al\mbox{.}(2024)]%
        {me_dhen}
\bibfield{author}{\bibinfo{person}{Xiaolong Liu}, \bibinfo{person}{Zhichen Zeng}, \bibinfo{person}{Xiaoyi Liu}, \bibinfo{person}{Siyang Yuan}, \bibinfo{person}{Weinan Song}, \bibinfo{person}{Mengyue Hang}, \bibinfo{person}{Yiqun Liu}, \bibinfo{person}{Chaofei Yang}, \bibinfo{person}{Donghyun Kim}, \bibinfo{person}{Wen-Yen Chen}, {et~al\mbox{.}}} \bibinfo{year}{2024}\natexlab{}.
\newblock \showarticletitle{A Collaborative Ensemble Framework for CTR Prediction}.
\newblock \bibinfo{journal}{\emph{arXiv preprint arXiv:2411.13700}} (\bibinfo{year}{2024}).
\newblock


\bibitem[Ma et~al\mbox{.}(2018b)]%
        {MMoE2018}
\bibfield{author}{\bibinfo{person}{Jiaqi Ma}, \bibinfo{person}{Zhe Zhao}, \bibinfo{person}{Xinyang Yi}, \bibinfo{person}{Jilin Chen}, \bibinfo{person}{Lichan Hong}, {and} \bibinfo{person}{Ed~H. Chi}.} \bibinfo{year}{2018}\natexlab{b}.
\newblock \showarticletitle{Modeling Task Relationships in Multi-task Learning with Multi-gate Mixture-of-Experts}. In \bibinfo{booktitle}{\emph{KDD}}. \bibinfo{publisher}{{ACM}}, \bibinfo{pages}{1930--1939}.
\newblock


\bibitem[Ma et~al\mbox{.}(2018a)]%
        {ESMM2018}
\bibfield{author}{\bibinfo{person}{Xiao Ma}, \bibinfo{person}{Liqin Zhao}, \bibinfo{person}{Guan Huang}, \bibinfo{person}{Zhi Wang}, \bibinfo{person}{Zelin Hu}, \bibinfo{person}{Xiaoqiang Zhu}, {and} \bibinfo{person}{Kun Gai}.} \bibinfo{year}{2018}\natexlab{a}.
\newblock \showarticletitle{Entire space multi-task model: An effective approach for estimating post-click conversion rate}. In \bibinfo{booktitle}{\emph{SIGIR}}. \bibinfo{pages}{1137--1140}.
\newblock


\bibitem[Mao et~al\mbox{.}(2023)]%
        {FinalMLP2023}
\bibfield{author}{\bibinfo{person}{Kelong Mao}, \bibinfo{person}{Jieming Zhu}, \bibinfo{person}{Liangcai Su}, \bibinfo{person}{Guohao Cai}, \bibinfo{person}{Yuru Li}, {and} \bibinfo{person}{Zhenhua Dong}.} \bibinfo{year}{2023}\natexlab{}.
\newblock \showarticletitle{FinalMLP: An Enhanced Two-Stream MLP Model for CTR Prediction}.
\newblock \bibinfo{journal}{\emph{arXiv preprint arXiv:2304.00902}} (\bibinfo{year}{2023}).
\newblock


\bibitem[Misra et~al\mbox{.}(2016)]%
        {Cross-Stitch}
\bibfield{author}{\bibinfo{person}{Ishan Misra}, \bibinfo{person}{Abhinav Shrivastava}, \bibinfo{person}{Abhinav Gupta}, {and} \bibinfo{person}{Martial Hebert}.} \bibinfo{year}{2016}\natexlab{}.
\newblock \showarticletitle{Cross-Stitch Networks for Multi-task Learning}. In \bibinfo{booktitle}{\emph{CVPR}}. \bibinfo{publisher}{{IEEE} Computer Society}, \bibinfo{pages}{3994--4003}.
\newblock


\bibitem[Pan et~al\mbox{.}(2018)]%
        {FwFM2018}
\bibfield{author}{\bibinfo{person}{Junwei Pan}, \bibinfo{person}{Jian Xu}, \bibinfo{person}{Alfonso~Lobos Ruiz}, \bibinfo{person}{Wenliang Zhao}, \bibinfo{person}{Shengjun Pan}, \bibinfo{person}{Yu Sun}, {and} \bibinfo{person}{Quan Lu}.} \bibinfo{year}{2018}\natexlab{}.
\newblock \showarticletitle{Field-weighted factorization machines for click-through rate prediction in display advertising}. In \bibinfo{booktitle}{\emph{Proceedings of the 2018 World Wide Web Conference (WWW)}}. \bibinfo{pages}{1349--1357}.
\newblock


\bibitem[Pan et~al\mbox{.}(2024)]%
        {pan2024ads}
\bibfield{author}{\bibinfo{person}{Junwei Pan}, \bibinfo{person}{Wei Xue}, \bibinfo{person}{Ximei Wang}, \bibinfo{person}{Haibin Yu}, \bibinfo{person}{Xun Liu}, \bibinfo{person}{Shijie Quan}, \bibinfo{person}{Xueming Qiu}, \bibinfo{person}{Dapeng Liu}, \bibinfo{person}{Lei Xiao}, {and} \bibinfo{person}{Jie Jiang}.} \bibinfo{year}{2024}\natexlab{}.
\newblock \showarticletitle{Ads recommendation in a collapsed and entangled world}. In \bibinfo{booktitle}{\emph{Proceedings of the 30th ACM SIGKDD Conference on Knowledge Discovery and Data Mining}}. \bibinfo{pages}{5566--5577}.
\newblock


\bibitem[Qu et~al\mbox{.}(2016)]%
        {PNN2016}
\bibfield{author}{\bibinfo{person}{Yanru Qu}, \bibinfo{person}{Han Cai}, \bibinfo{person}{Kan Ren}, \bibinfo{person}{Weinan Zhang}, \bibinfo{person}{Yong Yu}, \bibinfo{person}{Ying Wen}, {and} \bibinfo{person}{Jun Wang}.} \bibinfo{year}{2016}\natexlab{}.
\newblock \showarticletitle{Product-based neural networks for user response prediction}. In \bibinfo{booktitle}{\emph{2016 IEEE 16th International Conference on Data Mining (ICDM)}}. IEEE, \bibinfo{pages}{1149--1154}.
\newblock


\bibitem[Rendle(2010)]%
        {FM2010}
\bibfield{author}{\bibinfo{person}{Steffen Rendle}.} \bibinfo{year}{2010}\natexlab{}.
\newblock \showarticletitle{Factorization machines}. In \bibinfo{booktitle}{\emph{2010 IEEE International Conference on Data Mining (ICDM)}}. IEEE, \bibinfo{pages}{995--1000}.
\newblock


\bibitem[Shazeer et~al\mbox{.}(2017a)]%
        {shazeer2017outrageously}
\bibfield{author}{\bibinfo{person}{Noam Shazeer}, \bibinfo{person}{Azalia Mirhoseini}, \bibinfo{person}{Krzysztof Maziarz}, \bibinfo{person}{Andy Davis}, \bibinfo{person}{Quoc Le}, \bibinfo{person}{Geoffrey Hinton}, {and} \bibinfo{person}{Jeff Dean}.} \bibinfo{year}{2017}\natexlab{a}.
\newblock \showarticletitle{Outrageously large neural networks: The sparsely-gated mixture-of-experts layer}.
\newblock \bibinfo{journal}{\emph{arXiv preprint arXiv:1701.06538}} (\bibinfo{year}{2017}).
\newblock


\bibitem[Shazeer et~al\mbox{.}(2017b)]%
        {moe}
\bibfield{author}{\bibinfo{person}{Noam Shazeer}, \bibinfo{person}{Azalia Mirhoseini}, \bibinfo{person}{Krzysztof Maziarz}, \bibinfo{person}{Andy Davis}, \bibinfo{person}{Quoc~V. Le}, \bibinfo{person}{Geoffrey~E. Hinton}, {and} \bibinfo{person}{Jeff Dean}.} \bibinfo{year}{2017}\natexlab{b}.
\newblock \showarticletitle{Outrageously Large Neural Networks: The Sparsely-Gated Mixture-of-Experts Layer}. In \bibinfo{booktitle}{\emph{ICLR}}. \bibinfo{publisher}{OpenReview.net}.
\newblock


\bibitem[Song et~al\mbox{.}(2019)]%
        {AutoInt2019}
\bibfield{author}{\bibinfo{person}{Weiping Song}, \bibinfo{person}{Chence Shi}, \bibinfo{person}{Zhiping Xiao}, \bibinfo{person}{Zhijian Duan}, \bibinfo{person}{Yewen Xu}, \bibinfo{person}{Ming Zhang}, {and} \bibinfo{person}{Jian Tang}.} \bibinfo{year}{2019}\natexlab{}.
\newblock \showarticletitle{Autoint: Automatic feature interaction learning via self-attentive neural networks}. In \bibinfo{booktitle}{\emph{Proceedings of the 28th ACM International Conference on Information and Knowledge Management (CIKM)}}. \bibinfo{pages}{1161--1170}.
\newblock


\bibitem[Su et~al\mbox{.}(2024)]%
        {STEM2023}
\bibfield{author}{\bibinfo{person}{Liangcai Su}, \bibinfo{person}{Junwei Pan}, \bibinfo{person}{Ximei Wang}, \bibinfo{person}{Xi Xiao}, \bibinfo{person}{Shijie Quan}, \bibinfo{person}{Xihua Chen}, {and} \bibinfo{person}{Jie Jiang}.} \bibinfo{year}{2024}\natexlab{}.
\newblock \showarticletitle{STEM: Unleashing the Power of Embeddings for Multi-task Recommendation}. In \bibinfo{booktitle}{\emph{Proceedings of the AAAI Conference on Artificial Intelligence}}, Vol.~\bibinfo{volume}{38}. \bibinfo{pages}{9002--9010}.
\newblock


\bibitem[Sun et~al\mbox{.}(2021)]%
        {FmFM2021}
\bibfield{author}{\bibinfo{person}{Yang Sun}, \bibinfo{person}{Junwei Pan}, \bibinfo{person}{Alex Zhang}, {and} \bibinfo{person}{Aaron Flores}.} \bibinfo{year}{2021}\natexlab{}.
\newblock \showarticletitle{Fm2: Field-matrixed factorization machines for recommender systems}. In \bibinfo{booktitle}{\emph{Proceedings of the Web Conference 2021}}. \bibinfo{pages}{2828--2837}.
\newblock


\bibitem[Tang et~al\mbox{.}(2020)]%
        {PLE2020}
\bibfield{author}{\bibinfo{person}{Hongyan Tang}, \bibinfo{person}{Junning Liu}, \bibinfo{person}{Ming Zhao}, {and} \bibinfo{person}{Xudong Gong}.} \bibinfo{year}{2020}\natexlab{}.
\newblock \showarticletitle{Progressive Layered Extraction {(PLE):} {A} Novel Multi-Task Learning {(MTL)} Model for Personalized Recommendations}. In \bibinfo{booktitle}{\emph{RecSys}}. \bibinfo{publisher}{{ACM}}, \bibinfo{pages}{269--278}.
\newblock


\bibitem[Wang et~al\mbox{.}(2021)]%
        {DCNv22021}
\bibfield{author}{\bibinfo{person}{Ruoxi Wang}, \bibinfo{person}{Rakesh Shivanna}, \bibinfo{person}{Derek Cheng}, \bibinfo{person}{Sagar Jain}, \bibinfo{person}{Dong Lin}, \bibinfo{person}{Lichan Hong}, {and} \bibinfo{person}{Ed Chi}.} \bibinfo{year}{2021}\natexlab{}.
\newblock \showarticletitle{DCN-V2: Improved deep \& cross network and practical lessons for web-scale learning to rank systems}. In \bibinfo{booktitle}{\emph{Proceedings of the Web Conference (WWW)}}. \bibinfo{pages}{1785--1797}.
\newblock


\bibitem[Wu et~al\mbox{.}(2024)]%
        {llm_survey}
\bibfield{author}{\bibinfo{person}{Likang Wu}, \bibinfo{person}{Zhi Zheng}, \bibinfo{person}{Zhaopeng Qiu}, \bibinfo{person}{Hao Wang}, \bibinfo{person}{Hongchao Gu}, \bibinfo{person}{Tingjia Shen}, \bibinfo{person}{Chuan Qin}, \bibinfo{person}{Chen Zhu}, \bibinfo{person}{Hengshu Zhu}, \bibinfo{person}{Qi Liu}, {et~al\mbox{.}}} \bibinfo{year}{2024}\natexlab{}.
\newblock \showarticletitle{A survey on large language models for recommendation}.
\newblock \bibinfo{journal}{\emph{World Wide Web}} \bibinfo{volume}{27}, \bibinfo{number}{5} (\bibinfo{year}{2024}), \bibinfo{pages}{60}.
\newblock


\bibitem[Xiao et~al\mbox{.}(2017)]%
        {AFM2017}
\bibfield{author}{\bibinfo{person}{Jun Xiao}, \bibinfo{person}{Hao Ye}, \bibinfo{person}{Xiangnan He}, \bibinfo{person}{Hanwang Zhang}, \bibinfo{person}{Fei Wu}, {and} \bibinfo{person}{Tat-Seng Chua}.} \bibinfo{year}{2017}\natexlab{}.
\newblock \showarticletitle{Attentional factorization machines: Learning the weight of feature interactions via attention networks}.
\newblock \bibinfo{journal}{\emph{arXiv preprint arXiv:1708.04617}} (\bibinfo{year}{2017}).
\newblock


\bibitem[Zbontar et~al\mbox{.}(2021)]%
        {zbontar2021barlow}
\bibfield{author}{\bibinfo{person}{Jure Zbontar}, \bibinfo{person}{Li Jing}, \bibinfo{person}{Ishan Misra}, \bibinfo{person}{Yann LeCun}, {and} \bibinfo{person}{St{\'e}phane Deny}.} \bibinfo{year}{2021}\natexlab{}.
\newblock \showarticletitle{Barlow twins: Self-supervised learning via redundancy reduction}. In \bibinfo{booktitle}{\emph{International conference on machine learning}}. PMLR, \bibinfo{pages}{12310--12320}.
\newblock


\bibitem[Zhang et~al\mbox{.}(2024)]%
        {zhang2024wukong}
\bibfield{author}{\bibinfo{person}{Buyun Zhang}, \bibinfo{person}{Liang Luo}, \bibinfo{person}{Yuxin Chen}, \bibinfo{person}{Jade Nie}, \bibinfo{person}{Xi Liu}, \bibinfo{person}{Daifeng Guo}, \bibinfo{person}{Yanli Zhao}, \bibinfo{person}{Shen Li}, \bibinfo{person}{Yuchen Hao}, \bibinfo{person}{Yantao Yao}, {et~al\mbox{.}}} \bibinfo{year}{2024}\natexlab{}.
\newblock \showarticletitle{Wukong: Towards a Scaling Law for Large-Scale Recommendation}.
\newblock \bibinfo{journal}{\emph{arXiv preprint arXiv:2403.02545}} (\bibinfo{year}{2024}).
\newblock


\bibitem[Zhang et~al\mbox{.}(2022)]%
        {zhang2022dhen}
\bibfield{author}{\bibinfo{person}{Buyun Zhang}, \bibinfo{person}{Liang Luo}, \bibinfo{person}{Xi Liu}, \bibinfo{person}{Jay Li}, \bibinfo{person}{Zeliang Chen}, \bibinfo{person}{Weilin Zhang}, \bibinfo{person}{Xiaohan Wei}, \bibinfo{person}{Yuchen Hao}, \bibinfo{person}{Michael Tsang}, \bibinfo{person}{Wenjun Wang}, {et~al\mbox{.}}} \bibinfo{year}{2022}\natexlab{}.
\newblock \showarticletitle{DHEN: A deep and hierarchical ensemble network for large-scale click-through rate prediction}.
\newblock \bibinfo{journal}{\emph{arXiv preprint arXiv:2203.11014}} (\bibinfo{year}{2022}).
\newblock


\bibitem[Zhang et~al\mbox{.}(2019)]%
        {DLRSSurvey2019}
\bibfield{author}{\bibinfo{person}{Shuai Zhang}, \bibinfo{person}{Lina Yao}, \bibinfo{person}{Aixin Sun}, {and} \bibinfo{person}{Yi Tay}.} \bibinfo{year}{2019}\natexlab{}.
\newblock \showarticletitle{Deep learning based recommender system: A survey and new perspectives}.
\newblock \bibinfo{journal}{\emph{ACM computing surveys (CSUR)}} \bibinfo{volume}{52}, \bibinfo{number}{1} (\bibinfo{year}{2019}), \bibinfo{pages}{1--38}.
\newblock


\bibitem[Zhang et~al\mbox{.}(2021)]%
        {ctr_survey}
\bibfield{author}{\bibinfo{person}{Weinan Zhang}, \bibinfo{person}{Jiarui Qin}, \bibinfo{person}{Wei Guo}, \bibinfo{person}{Ruiming Tang}, {and} \bibinfo{person}{Xiuqiang He}.} \bibinfo{year}{2021}\natexlab{}.
\newblock \showarticletitle{Deep learning for click-through rate estimation}.
\newblock \bibinfo{journal}{\emph{arXiv preprint arXiv:2104.10584}} (\bibinfo{year}{2021}).
\newblock


\bibitem[Zheng et~al\mbox{.}(2022)]%
        {AutoAttention2022}
\bibfield{author}{\bibinfo{person}{Zuowu Zheng}, \bibinfo{person}{Xiaofeng Gao}, \bibinfo{person}{Junwei Pan}, \bibinfo{person}{Qi Luo}, \bibinfo{person}{Guihai Chen}, \bibinfo{person}{Dapeng Liu}, {and} \bibinfo{person}{Jie Jiang}.} \bibinfo{year}{2022}\natexlab{}.
\newblock \showarticletitle{Autoattention: automatic field pair selection for attention in user behavior modeling}. In \bibinfo{booktitle}{\emph{2022 IEEE International Conference on Data Mining (ICDM)}}. IEEE, \bibinfo{pages}{803--812}.
\newblock


\bibitem[Zhu et~al\mbox{.}(2022)]%
        {BARS2022}
\bibfield{author}{\bibinfo{person}{Jieming Zhu}, \bibinfo{person}{Quanyu Dai}, \bibinfo{person}{Liangcai Su}, \bibinfo{person}{Rong Ma}, \bibinfo{person}{Jinyang Liu}, \bibinfo{person}{Guohao Cai}, \bibinfo{person}{Xi Xiao}, {and} \bibinfo{person}{Rui Zhang}.} \bibinfo{year}{2022}\natexlab{}.
\newblock \showarticletitle{Bars: Towards open benchmarking for recommender systems}. In \bibinfo{booktitle}{\emph{Proceedings of the 45th International ACM SIGIR Conference on Research and Development in Information Retrieval}}. \bibinfo{pages}{2912--2923}.
\newblock


\bibitem[Zhu et~al\mbox{.}(2023)]%
        {zhu2023final}
\bibfield{author}{\bibinfo{person}{Jieming Zhu}, \bibinfo{person}{Qinglin Jia}, \bibinfo{person}{Guohao Cai}, \bibinfo{person}{Quanyu Dai}, \bibinfo{person}{Jingjie Li}, \bibinfo{person}{Zhenhua Dong}, \bibinfo{person}{Ruiming Tang}, {and} \bibinfo{person}{Rui Zhang}.} \bibinfo{year}{2023}\natexlab{}.
\newblock \showarticletitle{Final: Factorized interaction layer for ctr prediction}. In \bibinfo{booktitle}{\emph{Proceedings of the 46th International ACM SIGIR Conference on Research and Development in Information Retrieval}}. \bibinfo{pages}{2006--2010}.
\newblock


\bibitem[Zhu et~al\mbox{.}(2020)]%
        {FuxiCTR2020}
\bibfield{author}{\bibinfo{person}{Jieming Zhu}, \bibinfo{person}{Jinyang Liu}, \bibinfo{person}{Shuai Yang}, \bibinfo{person}{Qi Zhang}, {and} \bibinfo{person}{Xiuqiang He}.} \bibinfo{year}{2020}\natexlab{}.
\newblock \showarticletitle{Fuxictr: An open benchmark for click-through rate prediction}.
\newblock \bibinfo{journal}{\emph{arXiv preprint arXiv:2009.05794}} (\bibinfo{year}{2020}).
\newblock


\end{thebibliography}

\clearpage
\appendix
\section{Appendix}
\subsection{Details of Datasets} \label{dataset}
\begin{itemize}[leftmargin=*]
    \item Criteo: The Criteo dataset is widely recognized and has been extensively used as a benchmark in the Criteo Display Advertising Challenge. This dataset comprises 45 million samples, each of which includes 13 numerical fields and 26 categorical fields.
    \item Avazu: The Avazu dataset, on the other hand, was employed in the Avazu CTR prediction competition. This dataset consists of 40 million samples, with each sample containing 24 categorical fields.
\end{itemize}

\begin{table}[h]
    \centering
    \caption{Statistics of the two Datasets}
    \begin{tabular}{lcccc}
    \toprule
    \textbf{Dataset} & \textbf{\#Train Size} & \textbf{\#Valid Size} & \textbf{\#Test Size}& \textbf{\#Fields} \\
    \midrule
    Avazu & 32.3M &4.0M &4.0M &24 \\
    Criteo & 33.0M &8.3M &4.6M &39 \\
    \bottomrule
    \label{table:dataset}
    \end{tabular}
\end{table}

\subsection{Details of Heterogeneous Expert} \label{hetero_experts}
Here we present the details of adopted heterogeneous experts in our experiments.
\paragraph{CIN}
Compressed Interaction Network (CIN) is designed to capture high-order feature interactions in a compressed manner. Following the best practices outlined in FuxiCTR \cite{FuxiCTR2020, BARS2022}, we adopt CIN experts with a single hidden layer of 276 dimensions for the Avazu dataset and hidden units of [16,16] for the Criteo dataset.
% \begin{equation}
%     \mathbf{X}_{k} = f\left(\mathbf{X}_{k-1} \circ \mathbf{X}_{0}\right)
% \end{equation}
\paragraph{DCN}
Deep Cross Network(DCN) is a widely used feature interaction module in CTR prediction tasks. It introduces a cross network that is efficient in learning high-order explicit feature interactions. In this paper, we adopt the DCN experts as a 3-layers Crossnet.
% \begin{equation}
%     \mathbf{X}_{k+1} = \mathbf{X}_0 \odot (\mathbf{W}_k \mathbf{X}_k + \mathbf{b}_k)+\mathbf{X}_0
% \end{equation}
\paragraph{DNN}
A Deep Neural Network (DNN) is a multi-layer neural network that learns hierarchical representations of input features. In this paper, we adopt DNN experts with a 3-hidden-layer architecture. Each layer have 500 hidden dimensions.
% \begin{equation}
%     \mathbf{X}_{k+1} = \sigma(\mathbf{W}_k \mathbf{X}_k + \mathbf{b}_k)
% \end{equation}
\paragraph{FM}
Factorized Machine(FM) calculates the interactions between all feature embeddings using inner products. In this paper, we only utilize 2nd-order interactions.

\end{document}